%% file: main.tex
\documentclass[letterpaper,twocolumn,10pt]{article}
\usepackage{usenix}

\usepackage{graphicx}
\usepackage{lineno}
\usepackage{dcolumn}
\usepackage{bm}
\usepackage{outlines}
\usepackage[dvipsnames]{xcolor}
\usepackage{amsmath}
\usepackage{amsthm}
\usepackage{amsfonts}

\usepackage{ifthen}
\usepackage{ulem}
\usepackage{xspace}
\usepackage{paralist}
\usepackage{cleveref}
\crefname{figure}{Figure}{Figures}
\crefname{table}{Table}{Tables}
\crefformat{section}{\S#2#1#3}
\Crefformat{section}{\S#2#1#3}
\crefformat{subsection}{\S#2#1#3}
\Crefformat{subsection}{\S#2#1#3}
\crefformat{subsubsection}{\S#2#1#3}
\Crefformat{subsubsection}{\S#2#1#3}
\usepackage{enumitem}
\usepackage{subcaption}
\usepackage{float}
\usepackage{fnpct}
\usepackage{microtype}
\usepackage{xparse}
\usepackage{xstring}
\usepackage{booktabs}
\usepackage{makecell}

\microtypecontext{spacing=nonfrench} 

\ExplSyntaxOn
\NewDocumentCommand{\btexttt}{m}
 {
  \texttt{
    \tl_set:Nn \l_tmpa_tl {#1}
    \tl_map_function:NN \l_tmpa_tl \__insert_breakpoints:n
  }
 }

\cs_new:Nn \__insert_breakpoints:n
 {
  \str_if_in:nnTF {./-} {#1}
   {#1\hspace{0pt}}
   {#1}
 }
\ExplSyntaxOff

\newcommand{\hide}[1] 
{
\ifthenelse{\boolean{false}}{#1}{}
}
\newtheoremstyle{mystyle} 
  {.3\topsep}               
  {.3\topsep}               
  {\itshape}              
  {}                      
  {\bfseries}             
  {.}                     
  {.5em}                  
  {}                      

\theoremstyle{mystyle}

\newcommand{\ie}{i.e.,\xspace}

\newcommand{\cmark}{{\color{ForestGreen}\checkmark}}
\newcommand{\xmark}{{\color{red}\textbf{\texttimes}}}

\newcommand{\eat}[1]{}
\usepackage{algorithm, algorithmicx}
\usepackage[noend]{algpseudocode}
\usepackage{multirow}
\usepackage{subcaption}
\usepackage{pifont}
\usepackage{enumitem}
\usepackage{refcount}
\setlist[itemize]{itemsep=0pt, partopsep=0pt, parsep=1pt, topsep=1pt}
\setlist[enumerate]{itemsep=0pt, partopsep=0pt, parsep=1pt, topsep=1pt}

\usepackage{minted}
\setminted{
  python3=true,
  fontsize=\footnotesize,
  autogobble,
  xleftmargin=1em,
  numbersep=4pt,
  highlightcolor=lightgray,
}
\setmintedinline{fontsize=\small}
\newmintedfile[pythoncode]{python}{}
\newmintedfile[yamlcode]{yaml}{
    baselinestretch=1.2
}

\usepackage{booktabs} 
\usepackage{siunitx} 
\newcommand{\rawsys}{SkyNomad\xspace}

\newcommand{\sys}{\btexttt{\rawsys}\xspace}

\newcommand{\MyPara}[1]{\vspace{.1em}\noindent\textbf{#1}~}

\begin{document}


\title{\rawsys: On Using Multi-Region Spot Instances to Minimize AI Batch Job Cost}


\author{
{\rm Zhifei Li}$^{*\dagger}$,
{\rm Tian Xia}$^{*\dagger}$,
{\rm Ziming Mao}$^\dagger$,
{\rm Zihan Zhou}$^\P$,
{\rm Ethan J. Jackson}$^\dagger$,
{\rm Jamison Kerney}$^\dagger$,
{\rm Zhanghao Wu}$^\dagger$,\\
{\rm Pratik Mishra}$^\S$,
{\rm Yi Xu}$^\dagger$,
{\rm Yifan Qiao}$^\dagger$,
{\rm Scott Shenker}$^{\dagger,\diamond}$,
{\rm Ion Stoica}$^\dagger$\\[1ex]
$^\dagger$UC Berkeley \quad $^\P$Shanghai Jiao Tong University \quad $^\S$AMD \quad $^\diamond$ICSI
}


\date{}

\maketitle

{\let\thefootnote\relax\footnote{$^*$Equal Contributions.}}

\input{0_abstract}

\input{1_introduction}
\input{2_background_and_motivation}

\input{4_design}
\input{5_implementation}
\input{7_evaluation}
\input{9_related}
\input{8_discussion}
\input{10_conclusion}

\bibliographystyle{plain}
\bibliography{paper}

\end{document}

%% file: 0_abstract.tex
\begin{abstract}
AI batch jobs such as model training, inference pipelines, and data analytics require substantial GPU resources and often need to finish before a deadline. Spot instances offer $3\text{--}10\times$ lower cost than on-demand instances, but their unpredictable availability makes meeting deadlines difficult.
Existing systems either rely solely on spot instances and risk deadline violations, or operate in simplified single-region settings. These approaches overlook substantial spatial and temporal heterogeneity in spot availability, lifetimes, and prices.
We show that exploiting such heterogeneity to access more spot capacity is the key to reduce the job execution cost.

We present \sys, a multi-region scheduling system that maximizes spot usage and minimizes cost while guaranteeing deadlines. \sys uses lightweight probing to estimate availability, predicts spot lifetimes, accounts for migration cost, and unifies regional characteristics and deadline pressure into a monetary cost model that guides scheduling decisions. Our evaluation shows that \sys achieves 1.25--3.96$\times$ cost savings in real cloud deployments and performs within 10\% cost differences of an optimal policy in simulation, while consistently meeting deadlines.
\end{abstract}

%% file: 1_introduction.tex
\section{Introduction}

\begin{figure}
    \centering
    \includegraphics[width=\linewidth]{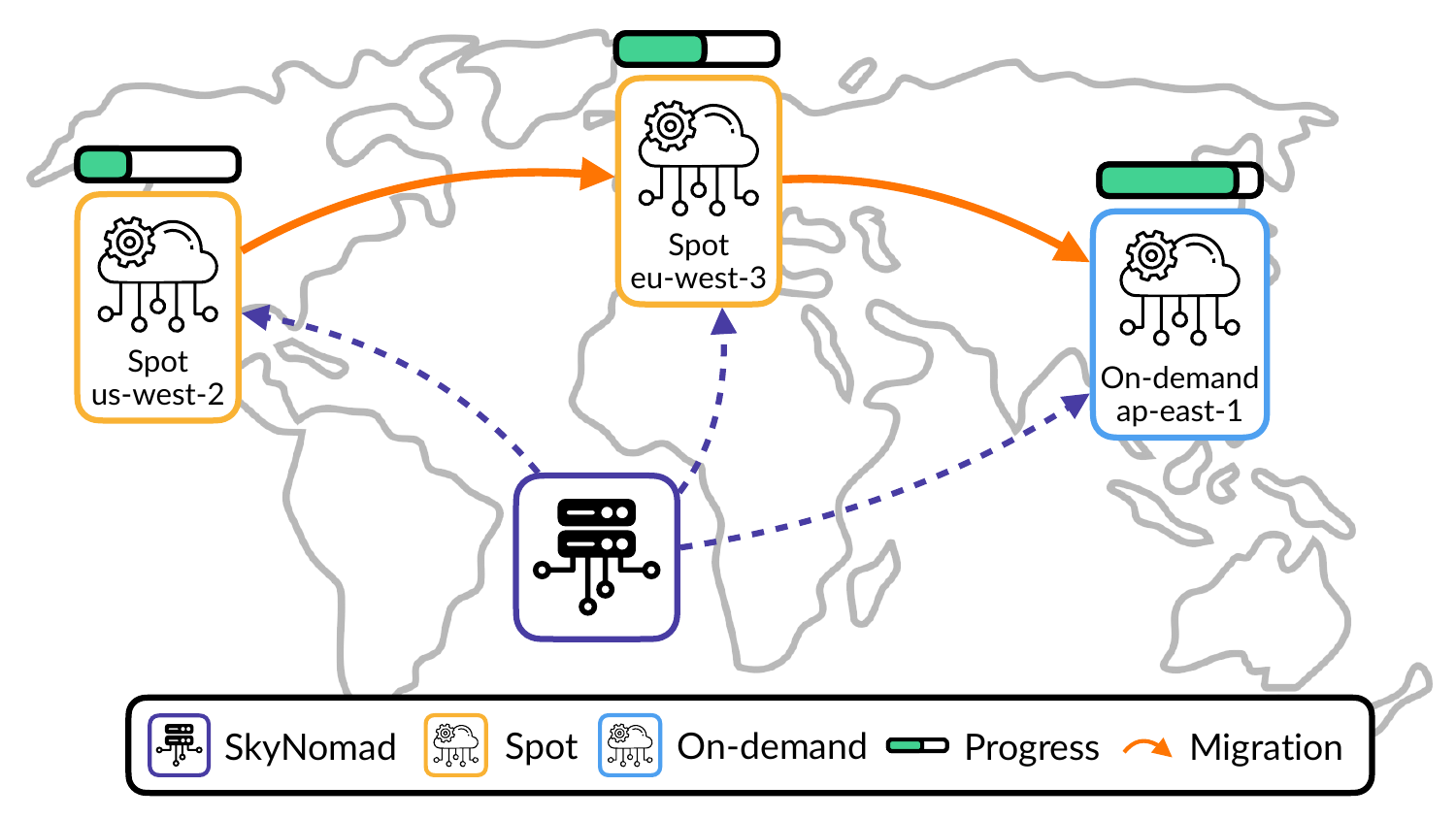}
    \caption{\sys Overview. \sys migrates AI batch jobs across multiple regions to harness available spot instances and minimize the cost. It monitors job progress and switches to on-demand instances when spot capacity is unavailable, ensuring that the job completes before its deadline.}
    \label{fig:sys-arch}
    \vspace{-1em}
\end{figure}

The emerging generative AI workloads demand unprecedented computational capacity. Large-scale model training, offline inference, and data analytics all require substantial GPU resources and often take hours or days to complete~\cite{brown2020languagemodelsfewshotlearners, hoffmann2022training, kaplan2020scalinglaws, chowdhery2023palm, touvron2023llama}. These long-running jobs, herein referred to as \emph{AI batch jobs}, often need to be completed by a given \emph{deadline}.
Such deadlines arise naturally and are workload-dependent, for example, periodic data processing must complete within a fixed time window (e.g., minutes to hours) to ensure that downstream analytics or dashboards reflect real-time data.

As model sizes and system complexity grow, these jobs become increasingly costly to operate. In this paper, we focus on minimizing the cost of running AI batch jobs while still meeting their deadlines. Many of these jobs assume they are interruptible and must tolerate failures (\cref{sec:chara-of-batch-jobs}), because such jobs often run for long periods, and GPU instances can experience failures at non-negligible rates~\cite{grattafiori2024llama3herdmodels, reiss2012heterogeneity}, the likelihood of a failure during execution is significant. To avoid wasting computation when failures occur, these jobs typically implement checkpoint-based recovery~\cite{gupta2024just, wan2025bytecheckpoint, wang2023gemini, narayanan2021efficient, rajbhandari2020zero, checkfreq, lazarus}.

To reduce cost, we can take advantage of a job's \emph{slack time}, which is the amount of delay it can tolerate before risking its deadline, and the preemptible nature of these jobs by using \emph{spot instances}.
Cloud providers offer these instances as a cheaper alternative to on-demand instances, typically 3--10$\times$ lower in cost~\cite{agmon2013deconstructing}.
However, they are subject to two limitations: preemption by the cloud provider and temporary unavailability in the spot market.
Therefore, spot capacity alone cannot guarantee that deadlines are met and must be used together with on-demand instances.
To maximize cost savings while still meeting the deadline, we need to consider \emph{where} and \emph{when} to use spot instances due to the heterogeneity they exhibit, spatially across regions and temporally across time.
When chosen properly, it is possible to access more spot capacity at lower prices and further reduce cost, especially for AI batch jobs that use GPUs and experience frequent preemptions~\cite{skyserve}. Existing work (\cref{sec:existing-work},~\cite{sagemaker2025,bamboo,parcae,cant-be-late,varuna,oobleck}) explores this opportunity but does not jointly consider the spatial and temporal dimensions of spot heterogeneity. Next, we describe these two forms of heterogeneity in detail.

\MyPara{Spatial heterogeneity across regions.} In public cloud providers, availability of spot instances can differ widely across cloud regions (\cref{sec:availability}). For example, \texttt{eu-central-1} may still have spot capacity while \texttt{us-east-1} has none. By leveraging spot instances across regions, a job can continue running on spot to make progress even when one region runs out of capacity, and benefiting from more cost savings. Pricing strategies also vary based on supply and usage patterns across regions, and can differ by up to 5$\times$~\cite{spot-location, aws-spot-pricing} (\cref{sec:pricing}), creating opportunities to proactively migrate to cheaper regions to save costs. However, combining these opportunities is challenging. Region availability is unknown ahead of time, and the scheduler must navigate multi-dimensional tradeoffs when availability and price diverge across regions, such as selecting between a high-cost region with high availability and a low-cost region with limited capacity. In addition, migrating jobs across regions with large checkpoints incurs non-trivial egress cost~\cite{AWS-CloudZero-Egress-2024,GCP-Network-Tiers-Pricing,Azure-Data-Transfer-Guide-2025}, so migration must be justified with improved availability or a lower price~\cite{tak2011move, khajeh2010cloud}.

\MyPara{Temporal heterogeneity across time.} Spot instance lifetimes, meaning the duration between provisioning and preemption, vary significantly across time (\cref{sec:lifetime}). We observe that short spot lifetimes, caused by frequent preemptions, often occur in short volatile periods. It is beneficial to avoid these periods and let the job use spot instances with longer lifetimes, which amortizes cold start costs (the cost to resume from the last checkpoint) and improves goodput. However, predicting spot lifetimes remains challenging.
Meanwhile, spot availability can change significantly over time (\cref{sec:availability}) and further complicate the scheduling. Early in the schedule, the policy can afford to wait for future periods with higher availability, running during those periods and pausing when unavailable to increase spot usage. As time passes and the deadline becomes tighter, the policy must become more conservative and prioritize steady progress using the resources available at the moment, trading off the possibility of better availability later against the need to ensure timely completion.

In summary, real cloud environments are complex and exhibit both spatial and temporal heterogeneity. Prior deadline-aware efforts~\cite{cant-be-late} explore this problem but consider only a single region, where many of these challenges do not arise. Extending to multiple regions is fundamentally harder because spatial heterogeneity must be leveraged while accounting for migration cost, and temporal heterogeneity must be evaluated across regions. A scheduler operating under these conditions must decide, at any point in time, whether to (i) use a spot instance in the current region or wait for one to appear, (ii) migrate to another region where spot capacity is available, or (iii) use an always-available on-demand instance to make progress to meet the deadline.

We design a multi-region spot scheduling policy to address the aforementioned challenges. First, we quantify the \textit{value} of the remaining progress in terms of monetary cost (\cref{sec:value-of-progress}). We estimate the expected cost of finishing the job on time given the current deadline pressure, which guides when the job can safely pause to wait for a better opportunity. Second, spot instances make progress only during their effective time, which is the uptime after cold start. To capture this, we use online regional availability probing (\cref{sec:probing}) to predict spot lifetimes (\cref{sec:duration-prediction}) and account for progress value only after the cold start, which helps avoid regions with volatile availability and frequent preemptions. Third, we compare this value against the operational cost in each region to guide the policy decisions (\cref{sec:unified-cost}), allowing the system to continuously evaluate regions and opportunistically migrate when a more cost-effective region appears (\cref{sec:putting-together}).

We implement this policy in \sys (\cref{fig:sys-arch}), an AI batch job execution system that draws spot capacity from multiple regions to complete jobs within deadlines. Users specify jobs with periodic checkpointing, and \sys migrates both computation and checkpoints across regions while following the policy's decisions to minimize cost.

To evaluate \sys, we deploy it on public clouds to run real batch jobs and experience real-time preemptions and migrations. As shown in \cref{sec:e2e}, \sys achieves 1.25--3.96$\times$ cost savings while meeting deadlines, compared to both prior research systems (e.g., Uniform Progress~\cite{cant-be-late}) and production systems (e.g., Amazon SageMaker~\cite{sagemaker2025}). We further conduct a comprehensive evaluation by replaying real preemption traces and comparing against existing policies, showing that \sys stays within 10\% of an omniscient policy across a wide range of configurations. These results demonstrate that, by leveraging deadline slack and drawing spot capacity from multiple regions, it is feasible to drastically reduce the cost of AI batch jobs.

In summary, this paper makes three contributions:

\begin{itemize}
    \item Identifying opportunities to leverage spot instances across regions to reduce the cost of AI batch jobs.
    \item The design of a scheduling policy that draws spot capacity from multiple regions while guaranteeing deadlines.
    \item \sys, an AI batch job execution system that runs jobs across regions with significant cost savings.
\end{itemize}

%% file: 2_background_and_motivation.tex
\section{Existing Systems for AI Batch Jobs}
\label{sec:existing-work}

In this section, we review several existing systems for running batch jobs in the cloud and discuss their limitations and missed opportunities for cost reduction (\cref{table:existing-system-comparison}). 

\subsection{On-Demand Only Systems}

The most common approach is to launch on-demand instances in the cloud, run the job to completion, and then terminate the instances. One example is AWS SageMaker~\cite{sagemaker2025}, a managed execution system for AI batch jobs (we discuss its managed spot support later). This approach guarantees timely execution before the deadline and avoids most failures, which simplifies system design. However, it is very costly.
Because on-demand execution ignores slack, jobs often finish well before their deadlines, incurring unnecessary cost.
\vspace{-1em}

\subsection{Spot Systems}

Prior work~\cite{sagemaker2025,parcae,bamboo,skyserve,mark,snape} uses spot to reduce cost. These systems fall into two categories: (1) batch-oriented spot-only systems and (2) online serving multi-region systems.

\MyPara{Spot-only systems fail to meet deadlines.} Several existing systems like SageMaker Managed Spot~\cite{sagemaker2025}, Parcae~\cite{parcae} and Bamboo~\cite{bamboo} rely exclusively on spot instances.
This works well when spot availability is high, but execution can be delayed indefinitely when spot capacity becomes scarce.
Prior work~\cite{skyserve} shows that, in the worst case, a region can run out of spot for 72 hours, during which the job makes no progress.

\MyPara{Online systems target a different workload.} Systems such as SkyServe~\cite{skyserve} run online LLM serving on spot instances across multiple regions and prioritize responsiveness, provisioning on-demand when spot is insufficient to maintain latency SLOs. Because they process requests immediately and cannot exploit slack, they miss cost-saving opportunities and are unsuitable for batch jobs.

\begin{table}[t]
\centering
\begin{tabular}{lcccc}
\toprule
& \makecell{Deadline\\Guarantee} 
& \makecell{Spot\\Instance} 
& \makecell{Multi-\\Region}  \\
\midrule
SageMaker~\cite{sagemaker2025}        & \cmark & \xmark & \xmark \\
Spot Only~\cite{sagemaker2025,bamboo,parcae} & \xmark & \cmark & \xmark \\
SkyServe~\cite{skyserve}  & \xmark & \cmark & \cmark \\
Uniform Progress~\cite{cant-be-late}   & \cmark & \cmark & \xmark \\
\sys      & \cmark & \cmark & \cmark \\
\bottomrule
\end{tabular}
\caption{Comparison of systems. Spot Only Systems: SageMaker Managed Spot~\cite{sagemaker2025}, Parcae~\cite{parcae}, and Bamboo~\cite{bamboo}.}
\label{table:existing-system-comparison}
\vspace{-1em}
\end{table}

\subsection{Deadline-Aware Systems}

Uniform Progress (UP)~\cite{cant-be-late} is a deadline-aware policy that uses on-demand and spot instances interchangeably. It uses spot when available, falls back to on-demand when capacity is scarce or slack is low, and spreads job progress evenly over time to meet the deadline. While simple, UP ignores temporal variation in spot availability and runs in a single region; when that region has no spot capacity, it must rely on on-demand, pushing cost close to that of always using on-demand (\cref{sec:e2e}).

UP assumes that spot availability is sufficiently stable within a single region to allow steady progress. 
Our analysis in \cref{sec:background} shows that real-world availability and lifetime are highly non-uniform across both regions and time, making single-region policies inherently inefficient.
Therefore, a more effective policy could leverage multiple regions to access additional spot capacity and further reduce cost~\cite{aws-spot-best-practices}.

\begin{figure*}[t]
    \centering
    \begin{subfigure}{0.64\textwidth}
        \centering
        \includegraphics[width=\linewidth]{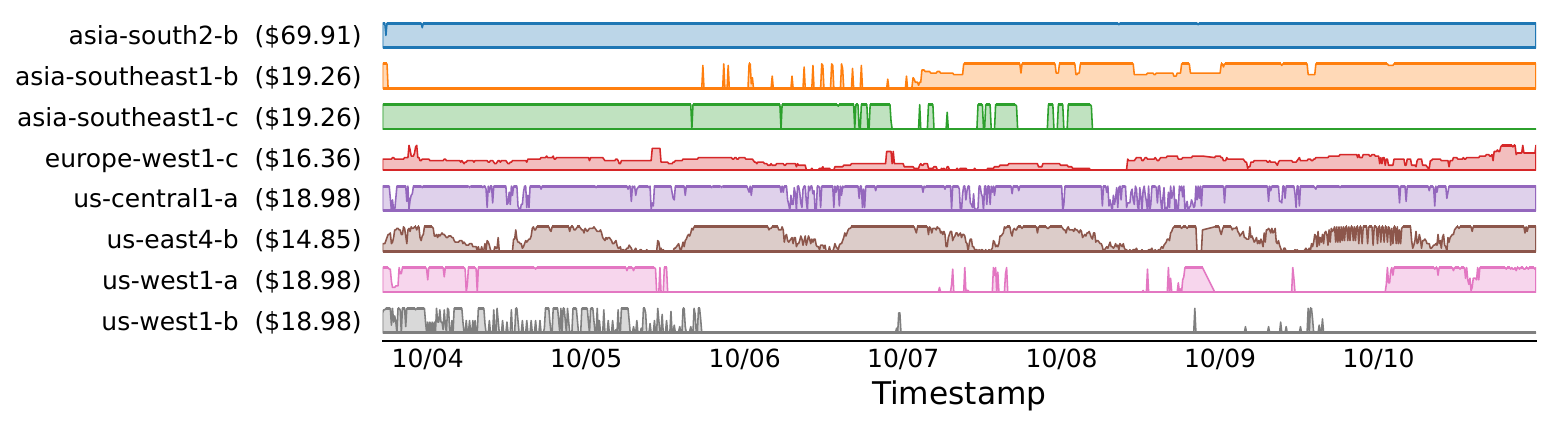}
        \label{fig:timeline-filled}
    \end{subfigure}
    \hfill
    \begin{subfigure}{0.33\textwidth}
        \centering
        \includegraphics[width=\linewidth]{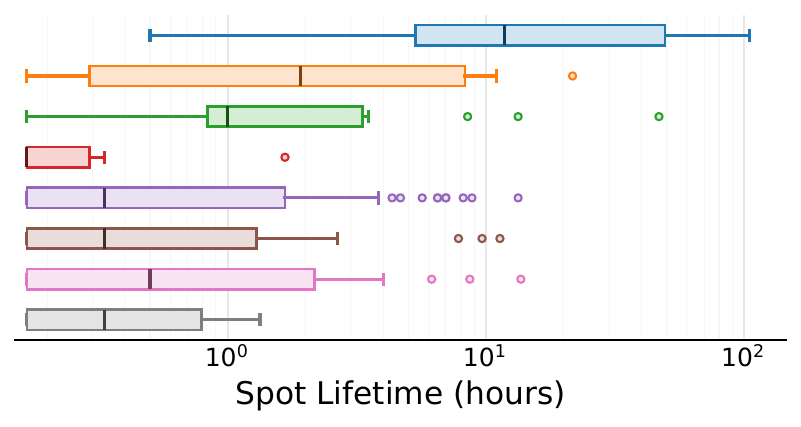}
        \label{fig:region-lifetime}
    \end{subfigure}
    \vspace{-2em}
    \caption{Spot availability and lifetime vary widely across regions. \textbf{Left}: availability for 16 \texttt{a3-highgpu-1g} (1$\times$H100) instances on GCP over 7 days, shown for 8 representative zones out of the 13 zones we collected. The y-axis shows how many instances are available at each moment. \textbf{Right}: box plot of the lifetime distribution, log-scaled. We consider a job that uses 16 instances in a gang-scheduled manner: the entire job is preempted if any single instance is preempted, and it resumes only when the full set of 16 instances becomes available again. We plot the resulting lifetimes under this scenario.
    }
    \label{fig:timeline}
\end{figure*}

\section{Background and Motivation}
\label{sec:background}

\subsection{Characteristics of AI Batch Jobs}
\label{sec:chara-of-batch-jobs}

\MyPara{Deadlines and slack time.} AI batch jobs are typically internal workloads that do not directly face end users. Unlike online serving, which prioritizes responsiveness and must meet tight service-level objectives (SLOs), AI batch jobs often run for long periods and do not require immediate response. Examples include model training and large-scale data analytics using AI models. Although they are not latency sensitive, such jobs usually have deadlines and cannot be delayed indefinitely. For instance, model training must finish before a scheduled release, and daily data pipelines must complete before the next cycle begins. In this paper, we assume jobs have an \textit{explicit} deadline specified in advance, and the job must finish before it. We also assume that the job’s execution time is shorter than the time available before the deadline, which creates \textit{slack time}, representing the amount of delay the job can tolerate before risking a deadline violation.

\MyPara{Interruptible.} Due to their long-running nature, batch jobs are generally designed to be interruptible. They must tolerate unexpected failures without losing significant progress. Training jobs typically perform periodic checkpointing, saving model parameters and optimizer states as recovery points~\cite{pytorch-fsdp, checkfreq}. If a failure occurs, the job resumes from the most recent checkpoint. Batch inference and data processing workflows can often be decomposed into independent units whose outputs are stored incrementally, with the processed data index serving as a lightweight checkpoint. This allows failures to be handled by restarting from unprocessed units rather than re-executing the entire dataset.

\MyPara{Cold start delay.} As model sizes grow, batch jobs incur a non-negligible cold start delay both at job launch and during every recovery. This delay includes instance provisioning, dependency installation, and downloading and loading model weights onto GPUs. For stateful jobs such as model training, it also includes transferring job state when recovering in a new location. For example, starting supervised fine-tuning for a 14B-parameter model on 8 GPUs can take 5--10 minutes. These cold start delays significantly affect overall progress and must be accounted for when designing scheduling policy.

\subsection{Heterogeneity of Spot Instances}
\label{sec:spot-heterogeneous}

Cloud providers offer a special class of instances, spot instances, as a cost-efficient option. They are typically $3$–$10\times$ cheaper~\cite{cant-be-late,skyserve,kumar2018survey} than regular on-demand instances but come with the risk of being \emph{preempted} at any time by the provider. Spot instances exhibit both spatial heterogeneity across regions and temporal heterogeneity over time, across multiple dimensions. Next, we describe these dimensions and the challenges in harnessing them to reduce cost.

\subsubsection{Non-Uniform Spot Availability Patterns}
\label{sec:availability}

Spot availability represents the periods during which a region has accessible spot capacity. An oracle scheduler with perfect future knowledge of availability could almost always run the job \emph{entirely} on spot instances, although not necessarily within the same region, achieving significant cost reductions. In practice, however, availability patterns vary across regions and fluctuate over time. When availability is unknown ahead of time, it becomes difficult to characterize each region and decide which one is the best region at any moment.

We collect spot availability traces for 16 \texttt{a3-highgpu-1g} instances (1$\times$H100) across 13 GCP regions. \cref{fig:timeline} shows 8 representative examples.
Different regions exhibit distinct availability patterns: some are generally available (\texttt{asia-south2-b}), some are mostly available but experience frequent preemptions (\texttt{us-central1-a}), and some are largely unavailable (\texttt{us-west1-b}).
We find that simultaneous preemptions across regions are rare, consistent with observations in~\cite{skyserve,spot-interruption-visible}. This makes regions complementary: at almost any given time, at least one region has available spot capacity. Prior work~\cite{skyserve} shows that the fraction of time in which spot instances are available quickly approaches 99\% once more than four regions are included, and we observed similar behavior for 6 regions in our trace.

Beyond cross-region heterogeneity, availability within a single region also exhibits large variance. For example, \texttt{asia-southeast1-c} is highly available in the first half of the trace but completely unavailable in the second half. \texttt{us-east4-b} shows a clear diurnal pattern, being available during nighttime and unavailable during daytime. These observations indicate that no single perfect region is consistently available (we discuss the limitation of \texttt{asia-south2-b} in \cref{sec:pricing}). As a result, a single-region policy is fundamentally limited, and handling non-uniform spot availability patterns becomes the central challenge.

\subsubsection{Highly Variable Spot Lifetimes}
\label{sec:lifetime}

Beyond availability, spot lifetime is another critical metric. It is defined as the duration for which a spot instance can be used before it is preempted. Because cold start time is non-negligible and does not contribute to job progress, short-lived spot instances provide little value if their lifetime is close to or smaller than the cold start delay. Let $d$ denote the cold start time. If a spot instance has a lifetime $t_0$, then the effective time during which the job progresses is $t_0 - d$. When $t_0$ approaches $d$, the job achieves very low goodput since most of the time is spent restarting rather than computing.

However, lifetime distributions are highly variable, across both regions and time. For example, \cref{fig:timeline} shows the distribution of spot lifetimes for 8 regions: median lifetimes range from under 1 hour (\texttt{europe-west1-c}) to over 20 hours (\texttt{asia-south2-b}), spanning more than an order of magnitude. Selecting a region with a long upcoming spot lifetime is essential because it reduces the number of recoveries and amortizes cold start overhead over longer execution periods.

We also observe that many spot preemptions occur within short time spans, which we refer to as \emph{volatile periods}. These periods produce many short-lived spot instances that become available briefly and are preempted almost immediately. For example, in region \texttt{us-central1-a}, 90\% of spot preemptions occur within 85 hours of a 15-day period.

\begin{figure}
    \centering
    \begin{subfigure}{0.48\linewidth}
        \centering
        \includegraphics[width=\linewidth]{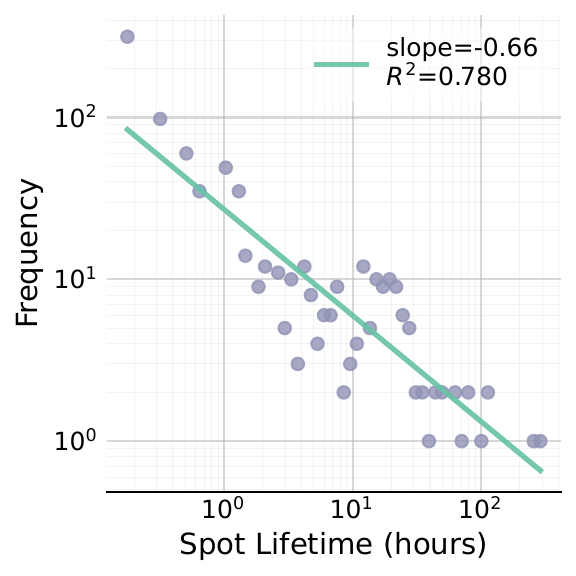}
        \caption{16$\times$H100 on GCP.}
        \label{fig:distribution-h100}
    \end{subfigure}
    \hfill
    \begin{subfigure}{0.48\linewidth}
        \centering
        \includegraphics[width=\linewidth]{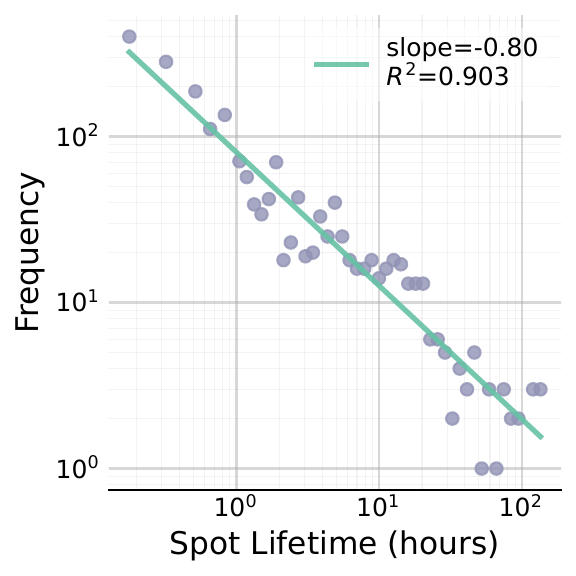}
        \caption{1$\times$V100 on AWS.}
        \label{fig:distribution-v100}
    \end{subfigure}
    \caption{
    Spot lifetime distributions on different accelerators.
    Scatter points show frequency counts in equal-width $\log_{10}$ bins, and solid lines are least-squares fits in log--log space, showing a clear heavy-tailed pattern~\cite{heavy-tail-ref,10.1145/166237.166255}.
    }
    \label{fig:distribution}
\end{figure}

Conversely, we observe that spot lifetimes follow a heavy-tailed distribution~\cite{kadupitiya2020modeling,spotlake}: the longer an instance has survived, the longer its expected remaining lifetime tends to be. As illustrated in \cref{fig:distribution}, the near-linear decay in log--log space (R$^2 \approx 0.78$--$0.90$) confirms this pattern: most instances are preempted within a few hours, yet a meaningful tail survives for tens of hours or more.

\subsubsection{Pricing Differences and Proactive Migration}
\label{sec:pricing}

\begin{figure}[t]
    \centering
    \begin{subfigure}[t]{0.54\columnwidth}
        \centering
        \includegraphics[width=\linewidth]{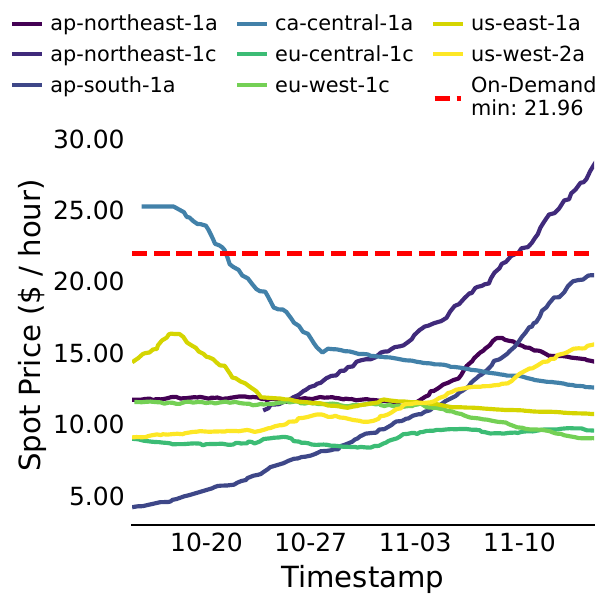}
        \caption{Spot prices.}
        \label{fig:price_variance}
    \end{subfigure}
    \hfill
    \begin{subfigure}[t]{0.432\columnwidth}
        \centering
        \includegraphics[width=\linewidth]{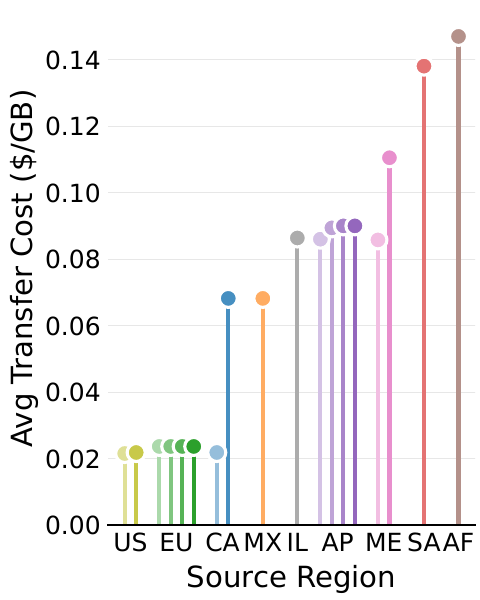}
        \caption{Egress costs.}
        \label{fig:egress}
    \end{subfigure}
    \caption{Cost differences across regions and time. (a) \textbf{Spot prices} for the same instance type. The dashed line shows the cheapest on-demand price. (b) \textbf{Egress costs} determined by the source region and ranging from \$0.02/GB (US, EU) to \$0.14/GB (SA, AF). Abbreviations follow AWS region naming conventions (e.g., AF for Africa, SA for South America).}
    \vspace{-1em}
    \label{fig:prices-all}
\end{figure}

Another form of heterogeneity resides in the pricing strategies. As shown in \Cref{fig:price_variance}, prices for the same instance type can differ significantly across regions, reaching up to 5$\times$. In most clouds (including AWS~\cite{aws-spot-pricing} and GCP~\cite{gcp-spot-pricing}), prices also fluctuate over time and can vary by up to 1.7$\times$ within only 12 days. Prior work \cite{cant-be-late, skyserve} often overlooks these differences and assumes static pricing, which oversimplifies the problem.

Beyond instance prices, cloud providers impose additional charges for cross-region data transfers. For jobs that maintain large checkpoints, such as training workloads, the checkpoint must be moved whenever the job migrates to a new region. This movement incurs non-negligible \emph{egress} costs~\cite{AWS-CloudZero-Egress-2024,GCP-Network-Tiers-Pricing,Azure-Data-Transfer-Guide-2025}, which vary depending on the source region. As shown in \Cref{fig:egress}, egress costs can differ by up to 7$\times$ across regions.
For example, a 14B-parameter fine-tuning job may need to move about 500GB of replicated checkpoints; this costs roughly \$10 from US to Europe but around \$40 from Asia to US.

Migrate to a region with a cheaper spot price may sound appealing, even when the current spot instance has not been preempted. This form of \emph{price-aware proactive migration}, however, is only beneficial when the migration cost is justified by a sufficiently long spot lifetime in the cheaper region. Let the price in region $i$ be $p_i$, the migration cost between regions $i$ and $j$ be $C_{i,j}$, the expected spot lifetime in region $i$ be $t_i$, and $d$ be the cold start delay. Suppose the job is currently running on a spot instance in region $a$. When a cheaper region $b$ becomes available, migrating is beneficial only if $C_{a,b} \le \frac{p_a - p_b}{t_b-d}$.

However, a single region rarely offers the best availability, the longest lifetime, and the lowest price simultaneously. For example, in \cref{fig:timeline-filled}, while mostly available, \texttt{asia-south2-b} is 4$\times$ more expensive than the cheapest region and approaches the on-demand price. In contrast, \texttt{us-east4-b} offers the lowest cost but experiences frequent preemptions and volatile periods. Along with migration cost, trading off among these four metrics creates a vast combinatorial decision space.

%% file: 4_design.tex
\section{Design}

\begin{figure}
	\centering
	\includegraphics[width=\linewidth]{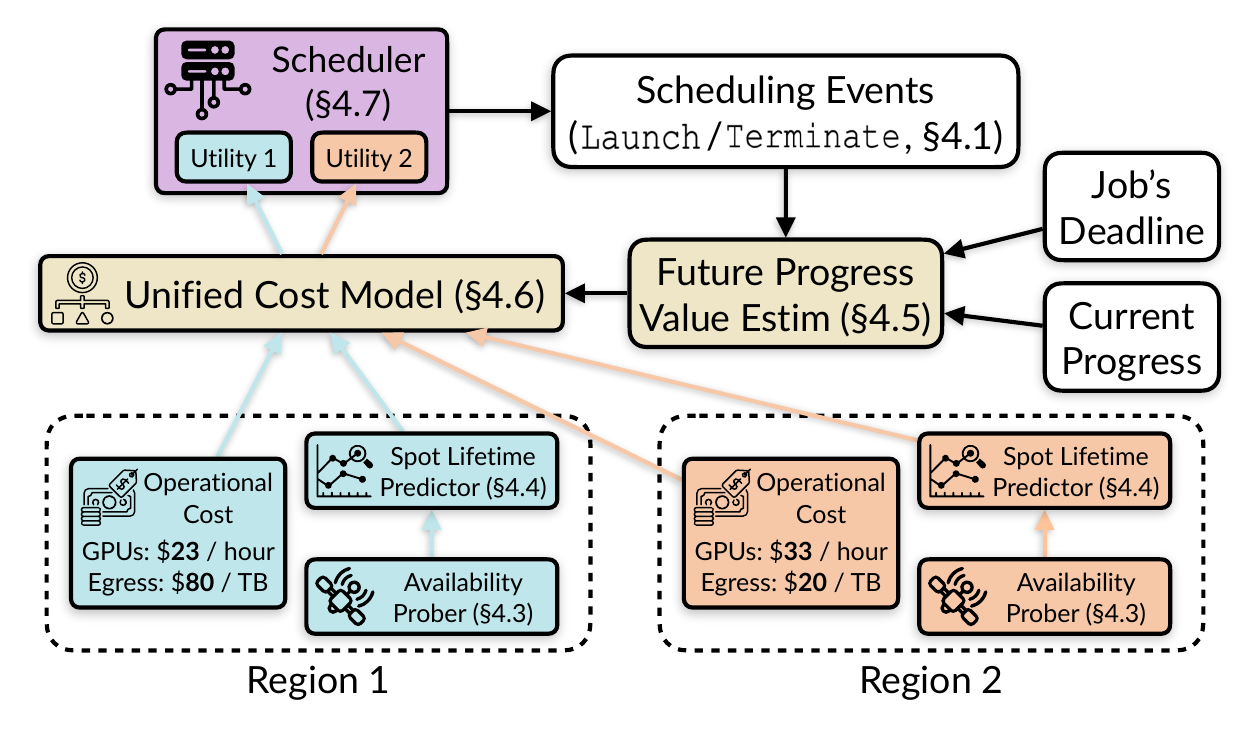}
	\caption{Design overview of \sys.}
	\label{fig:overview}
\end{figure}

We now present the design of \sys (\Cref{fig:overview}, \Cref{alg:policy}) to fully harness spot instances for AI batch jobs by accounting for spatial and temporal heterogeneity in availability, lifetime, and pricing (\cref{sec:spot-heterogeneous}). To stay aware of this heterogeneity, \sys continuously collects real-time information from each region (\cref{sec:probing}) and uses it together with survival analysis~\cite{aalen1978nonparametric} to predict spot lifetimes (\cref{sec:duration-prediction}). To decide whether current spot availability is worth taking, \sys considers the current deadline pressure and estimates the value (i.e, expected monetary cost) of the remaining progress (\cref{sec:value-of-progress}), which it uses as a reference when choosing to take or wait for a better opportunity. Finally, \sys applies a unified cost model (\cref{sec:unified-cost}) to compute the utility of each region under the current deadline pressure and proactively migrates to a better option when available, or pause when the slack time is sufficient for better future opportunities (\cref{sec:putting-together}).

\subsection{Problem Formulation}
\label{sec:problem-formulation}

We consider a single batch job, such as model training, that runs on a fixed group of instances at any time, with all instances in the group placed in the same region to benefit from high-bandwidth networking. This contrasts with autoscaling designs that distribute work across a dynamic number of instances and multiple regions simultaneously (we discuss this further in \cref{sec:autoscaling}).
We assume all instances in the group are gang-scheduled, and the preemption of any instance is treated as preemption of the entire group, without assuming any specific partial recovery strategy, which maximizes compatibility. Without loss of generality, we focus on the single-instance case in the remainder of the paper, as gang-scheduling allows instance groups to be treated as atomic units.

The job checkpoints periodically, enabling it to pause and resume in a different region after an interruption. We assume the execution time of the job can be estimated \emph{a priori} and denote it as $P$ units of work. At any time, the job may either remain idle, using slack time to wait for a better opportunity, or run on an instance. Once it starts or resumes on a new instance, it incurs a cold start delay of $d$ time units (\cref{sec:chara-of-batch-jobs}) before becoming ready to execute the job. Let $p(t)$ denote the total progress made by time $t$. The rate at which the job makes progress depends on the system state:
\begin{equation}
	\frac{\mathrm{d}p}{\mathrm{d}t} = \begin{cases}
		1 & \text{Ready}                          \\
		0 & \text{Otherwise (Cold Start or Idle)}
	\end{cases}
\end{equation}

The job must complete before a deadline $T$, that is, it must satisfy $p(T) \geq P$. Our goal is to minimize the total cost $\mathcal{C}$ while meeting the deadline, namely $\arg \min \mathcal{C}$ s.t. $p(T) \geq P$.

Let $\mathcal{R}$ denote the set of all regions, each offering both spot and on-demand instances. The job can run in any region in $\mathcal{R}$. At any time $t$, the system is in some state $s = (r,m)$, where $r \in \mathcal{R}$ is the current job location and $m \in \{\texttt{idle}, \texttt{spot}, \texttt{od}\}$ is the current mode. Mode \texttt{idle} means the job is waiting with no instance running; \texttt{spot} and \texttt{od} mean the job is running on a spot or on-demand instance, respectively. Spot instances are cheaper but may be preempted, whereas on-demand instances are always available but more expensive. The price for state $s=(r,m)$ at time $t$ is denoted $C_{(r,m)}(t)$.

State changes occur through three events:
\begin{itemize}
    \item \texttt{Launch}$(r,m)$: Attempts to launch an instance in region $r$ with mode $m \in \{\texttt{spot}, \texttt{od}\}$. This action always succeeds for $m = \texttt{od}$, and succeeds for $m = \texttt{spot}$ only if spot capacity is available, which is unknown until the attempt. A successful launch migrates the checkpoint to $r$ if it is not already there and sets the current state to $(r,m)$.
    \item \texttt{Terminate}: Stops the current instance, setting $m \gets \texttt{idle}$ while keeping $r$ unchanged, indicating that the checkpoint remains in the same region.
    \item \texttt{Preemption}: An external event that the cloud reclaims the spot instance, with the same effect as \texttt{Terminate}.
\end{itemize}

Given the above events, the system follows a trajectory that yields a total cost $\mathcal{C}$, consisting of two parts: compute cost and migration cost as the job moves between regions. Compute cost accumulates while running, at the price of the current region: $\mathcal{C}_\text{compute} = \int_{m \neq \texttt{idle}} C_{(r,m)}(t) \, \mathrm{d}t$.

Migration cost is incurred each time $r$ changes, i.e., when launching in a new region. Each such move requires transferring the checkpoint, costing $E_{r_i \to r_j} = e_{r_i \to r_j} \cdot S_\text{ckpt}$, where $e_{r_i \to r_j}$ is the per-byte egress cost and $S_\text{ckpt}$ is the checkpoint size (we assume $e_{r_i \to r_i}=0$). If the job migrates through regions $r_0 \to r_1 \to \cdots \to r_k$, the total migration cost is $\mathcal{C}_\text{migrate} = \sum_{i=1}^{k} E_{r_{i-1} \to r_i}$.
The total cost is thus $\mathcal{C} = \mathcal{C}_\text{compute} + \mathcal{C}_\text{migrate}$.

\subsection{Deadline-Aware Rules}
\label{sec:deadline-awareness}

First, we establish several rules
to ensure completion before the deadline without waste. These rules are adapted from~\cite{cant-be-late} with a multi-region extension.

\MyPara{Thrifty Rule.} The job should remain idle once all work is completed. That is, when $p(t) \geq P$, it should be in a state $s = (r, m)$ with $m = \texttt{idle}$.

\MyPara{Safety Net Rule.} When the job is idle and the remaining slack time is insufficient, it should switch to on-demand and stay on it until completion. Specifically, at time $t$, if $T - t < P - p(t) + 2 d$,
the scheduler switch to on-demand to guarantee the deadline. The term $2 d$ accounts for the cold start of the current launch and a potential fallback after a preemption.

\MyPara{Multi-Region Fallback.} In a multi-region setting where on-demand prices vary across regions, the fallback should choose the cheapest on-demand option when possible. Suppose the job is currently in region $r_0$ when the safety net is triggered. We select the fallback region $r$ as
\begin{equation}
\arg \min_{r \in \mathcal{R}} \left[ C_{(r,\texttt{od})}(t)\cdot (P - p(t) + d) + E_{r_0 \to r} \right]
\end{equation}
where the first term captures the on-demand cost in region $r$ to finish the remaining work and the second term captures the migration cost to move the checkpoint from $r_0$ to $r$.

\subsection{Spot Availability Probing}
\label{sec:probing}

Given the heterogeneity of spot availability (\cref{sec:availability}), we use online availability probing to characterize each candidate region and guide scheduling decisions. A \emph{probe} is a launch request that starts an instance with the same GPU configuration and immediately terminates it if the launch succeeds, providing a cost-efficient way to measure regional availability. We periodically send probes to candidate regions, with the interval set to two hours based on observation showing that availability patterns remain stable within this window (\Cref{{fig:timeline}}).
Unlike static historical traces that can become stale, probing provides real-time information during job execution.

\MyPara{Virtual Instance.} Based on probe results, we maintain a view of \emph{virtual instances}, as if an instance were continuously running in each region and receiving real-time preemptions. This is motivated by the observation that allocation failures (the inability to launch new spot) are strongly correlated with preemption events~\cite{cant-be-late}. For each region $r$, we build such a virtual instance view from observations of the form $(t_i, o_i)$, where $t_i$ is a timestamp and $o_i \in {0, 1}$ indicates availability. Observations arise from four sources: (1) probes (success $o=1$, failure $o=0$), (2) \texttt{Launch} operations (success $o=1$, failure $o=0$), (3) preemption events ($o=0$), and (4) \texttt{Terminate} operations when we proactively migrate away from a region ($o=0$). This logical construct lets us infer when an instance \emph{would} have been preempted without actually running it: whenever availability changes from $1$ to $0$, that is, $o_{i-1} = 1$ and $o_i = 0$. We treat such an event at time $t_i$ as a preemption of the virtual instance. Similarly, when availability changes from $0$ to $1$, we treat it as a provisioning of the virtual instance.

\subsection{Spot Lifetime Prediction}
\label{sec:duration-prediction}

As discussed in \cref{sec:lifetime}, knowing the future spot lifetime in each region helps the scheduler favor longer-lived spot instances, reducing the number of preemptions and amortizing both cold start overhead and migration cost. We predict this lifetime using the virtual instance trace in each region and, to incorporate the heavy-tailed distribution, condition our prediction on the current \emph{age} of the virtual instance, defined as the time since its last preemption. We apply survival analysis to estimate the lifetime distribution, and further refine the model to account for sudden bursts of preemptions based on the observation of volatile periods.

\subsubsection{Survival Analysis}

We first extract the historical lifetimes of the virtual instance, defined as each continuous period of availability. Each lifetime begins when a probe or launch succeeds ((1) and (2) in \cref{sec:probing}) and ends either with an observed preemption (3) or is right-censored when we proactively migrate away from the region (4). This yields a set of virtual instance lifetimes. We use $e(l)$ to denote the number of preemptions that occur at lifetime $l$ (end with (3)), and $c(l)$ to denote the number of proactive migrations that occur at lifetime $l$ (end with (4)).

Based on recent observations of the virtual instance, we first compute its age, denoted by $a(t)$. For example, if at time $t$ the last three probes succeeded and the fourth most recent failed, with a probe interval of two hours, then $a(t) = 6$ hours. Given this age, our goal is to predict how long the region will remain available. This requires estimating the conditional expected remaining lifetime $\mathbb{E}[L - a(t) \mid L > a(t)]$, where $L$ is the random variable for lifetime. Directly estimating is complicated by censored data, so we adopt survival analysis.

The key quantity in survival analysis is the \emph{hazard rate} $h(l)$, which is the instantaneous risk of termination at lifetime $l$ given survival up to that point. We estimate $h(l)$ using the Nelson-Aalen estimator~\cite{aalen1978nonparametric}:
\begin{equation}
	h(l) = \frac{e(l)}{n(l)}\text{ with }n(l)=\sum_{x \geq l}(e(x)+c(x))
\end{equation}
where $e(l)$ is the number of preemptions observed at lifetime $l$, $c(l)$ is the number of right-censored observations (proactive migrations) at lifetime $l$, and $n(l)$ is the at-risk population, i.e., instances with lifetime at least $l$. Proactively migrated instances are right-censored: they contribute to $n(l)$ but not $e(l)$. This estimator is \emph{non-parametric}, meaning it makes no distributional assumptions, and it naturally captures the heavy-tailed pattern: when long-lived instances exhibit lower preemption risk, fewer preemptions occur at larger $l$, so $e(l)$ decreases and $h(l)$ decreases with $l$.

Summing $h(l)$ yields the \emph{cumulative hazard} $H(l) = \sum_{l_i \leq l} h(l_i)$, from which we derive the \emph{survival function} $S(l) = \exp(-H(l)) = \Pr(L > l)$.
For an instance that has already survived to age $a(t)$, its expected remaining lifetime is:
\begin{equation}
    \label{eq:lifetime-prediction}
	\bar{L}(a(t)) = \mathbb{E}[L - a(t) \mid L > a(t)] = \frac{1}{S(a(t))} \sum_{l_i > a(t)} S(l_i)
\end{equation}

\subsubsection{Volatility and Risk Assessment}

To ensure steady progress, the scheduler must promptly detect volatile periods (\cref{sec:lifetime}) and avoid placing jobs in regions experiencing frequent preemptions. We achieve this by incorporating volatility awareness into lifetime estimation and derive an adjusted survival function $\tilde{S}(l)$. We use $\tilde{S}$ in place of $S$ when computing $\bar{L}(a(t))$, penalizing regions that are currently in such periods. The key idea is to compare observed preemptions against what the long-term hazard rate $h(l)$ would predict: if a region experiences far more preemptions than expected, it is likely in a volatile period. For a time window $W$ ending at the current time, we define the \emph{volatility ratio} as $\gamma_W = \frac{e_W}{\sum_{t \in W} h(a(t))}$,
where $e_W$ is the observed number of preemptions in $W$, and the denominator is the expected number, obtained by summing the hazard rate $h$ at each observation time in the window. When $\gamma_W > 1$, more preemptions occurred than expected, indicating a volatile period.

We take $\gamma^* = \max_{W=(t_0\text{ to }t),\,t_0\in[0,t)} \gamma_W$ over all window lengths ending at the current time to conservatively capture the most severe preemption density at any time scale. We then adjust the survival curve by scaling the cumulative hazard: using $\tilde{S}(l) = \exp(\gamma^* \cdot -H(l))$ to replace $S(l)$ in \cref{eq:lifetime-prediction}.

\subsection{Future Progress Value Estimation}
\label{sec:value-of-progress}

With characterized spot behavior in every region, the question now becomes whether the current spot opportunity is worth taking or if it is better to pause and wait for a future one. To evaluate this, we establish the monetary value of future progress: by comparing the \emph{expected} cost of making the same amount of progress in the future with the \emph{observed} spot (or on-demand) operational cost at the moment, we can decide whether the system should use the current opportunity or wait. This evaluation should depend on the \emph{deadline pressure} at time $t$, defined as the remaining progress to make divided by the remaining time before the deadline:
\begin{equation}
    \theta(t) = \frac{P - p(t)}{T - t}
\end{equation}
as well as the \emph{average progress} so far:
\begin{equation}
	\tilde{\theta}(t) = \frac{p(t)}{t}
\end{equation}

\MyPara{Design Principles.} We first propose several general principles for evaluating the value of any progress:

\begin{itemize}
    \item \textbf{Equilibrium anchoring}: When the job has progressed as expected ($\theta(t) = \tilde{\theta}(t)=\frac{P}{T}$), the policy should accept any available spot instances but avoid using on-demand instances. In this case, historical spot supply has been sufficient to meet the deadline, so if only on-demand capacity is available at the moment, it is better to wait rather than spend extra on it.
    \item \textbf{Monotonicity}: Higher deadline pressure $\theta(t)$ should yield a higher value, since with less slack the policy must make progress to avoid risking a deadline violation.
    \item \textbf{Scale invariance}: The evaluation should be consistent regardless of the absolute values of $P$ and $T$, and depend only on their ratio.
\end{itemize}

\MyPara{Value of Progress.}
Based on those principles, we estimate this value by directly comparing the deadline pressure with the average progress so far:
\begin{equation}
	\label{eq:value-of-progress}
	V(t) = C_{\texttt{od}} \cdot \frac{\theta(t)}{\tilde{\theta}(t)}
\end{equation}
where $C_{\texttt{od}} = \min_r C_{(r,\texttt{od})}$ is the cheapest on-demand price across all regions (we assume on-demand prices do not change across time). It satisfies all three principles:

\begin{itemize}
    \item \textbf{Equilibrium anchoring}: When $\theta(t) = \tilde{\theta}(t)$, the progress is evaluated as $C_{\texttt{od}}$. We assume all spot instances are cheaper than on-demand instances for simplicity, so the policy will accept any available spot instances, as they make progress at lower cost, while avoiding all on-demand instances, as $C_{\texttt{od}}$ is no greater than any on-demand price and thus does not justify using them.
    \item \textbf{Monotonicity}: Fix $t$. Then $\theta(t)$ decreases with $p(t)$ and $\tilde{\theta}(t)$ increases with $p(t)$. Therefore, higher $\theta(t)$ yields a higher ratio $\frac{\theta(t)}{\tilde{\theta}(t)}$, hence a higher $V(t)$.
    \item \textbf{Scale invariance}: $V(t)$ depends only on the ratio between progress and time, through $\theta(t)$ and $\tilde{\theta}(t)$, and not on the absolute values of $P$ or $T$.
\end{itemize}

\subsection{Unified Cost Model}
\label{sec:unified-cost}

With the value of progress $V(t)$ established (\cref{sec:value-of-progress}), we evaluate each candidate state $s = (r,m) \in \mathcal{R} \times \{\texttt{spot}\}$ by computing its {expected net utility} (we discuss the \texttt{od} and \texttt{idle} cases in the following paragraph). Let the current state be $s_0 = (r_0, m_0)$, and let $\bar{L}_{s}$ denote the expected remaining lifetime (\cref{sec:duration-prediction}). Over the instance's lifetime, we define
\begin{equation}
    \bar{L}_{s} \cdot U_{s} = V(t) \cdot \max(0, \bar{L}_{s} - d) - C_{(r,m)}(t) \cdot \bar{L}_{s} - E_{r_0\to r}
\end{equation}
where $U_{s}$ denotes the utility per unit time over the lifetime. The three terms on the right-hand side capture: (1) the value of progress made during effective runtime, excluding the cold start time $d$; (2) the total compute cost $C_{(r,m)}(t)$ (for simplicity, we assume the price does not change within the instance's lifetime); and (3) the one-time migration cost $E_{r_0 \to r}$ if the target region differs from the current checkpoint location.

For comparison across candidates, we normalize by the expected lifetime to obtain the utility $U_s$:
\begin{equation}
	\label{eq:utility}
	U_{s} = \underbrace{V(t) \cdot \eta_{s}}_{\substack{\text{Expected Value of}\\\text{Effective Progress}}} - \underbrace{C_{(r,m)}(t)}_{\substack{\text{Operational}\\\text{Compute Cost}}} - \underbrace{\frac{E_{r_0\to r}}{\bar{L}_{s}}}_{\substack{\text{Amortized}\\\text{Migration Cost}}}
\end{equation}
where $\eta_{s} = \max(0, \bar{L}_{s} - d) / \bar{L}_{s}$ is the \emph{effectiveness} of candidate state $s$, i.e., the fraction of its lifetime spent doing useful work. $V(t) \cdot \eta_{s}$ thus denotes the effective value of the progress. Short-lived spot instances have lower effectiveness, since cold-start overhead dominates their runtime.
The second and third terms capture the operational cost of running in $s$, namely the compute cost combined with the amortized egress cost. The migration cost $E_{r_0 \to r}$ is amortized over the lifetime, so longer-lived instances amortize this fixed cost more effectively.

\MyPara{Special cases.} For on-demand instances, we assume $\bar{L}_{(r,\texttt{od})} \to \infty$ (no preemption), so $\eta_{(r,\texttt{od})} \to 1$ and the migration cost is fully amortized, giving $U_{(r,\texttt{od})} = V(t) - C_{(r,\texttt{od})}$. For idling, no cost is incurred but no progress is made, so $U_{(r,\texttt{idle})} = 0$. This means running on $s$ is worthwhile only when $U > 0$, \ie when the effective value of progress exceeds its cost.

\subsection{Putting it All Together}
\label{sec:putting-together}

\begin{algorithm}[t]
	\caption{\sys Scheduling Policy}
	\label{alg:policy}
	\begin{algorithmic}[1]
		\State \textbf{Input:} Job requiring total progress $P$, deadline $T$, and candidate regions $\mathcal{R}$
		\State $(r_0,m_0) \gets($initial region, \texttt{idle}$)$
		\While{$p(t) < P$} \Comment{Every scheduling step}
		\If{\Call{SafetyNet}{t}} \Comment{\cref{sec:deadline-awareness}} \label{alg:line:safety}
		\State \texttt{\textbf{Launch}} \texttt{od} if $m \neq \texttt{od}$; \textbf{continue} \label{alg:line:safety-end}
		\EndIf
		\State \Call{ProbeRegions}{$\mathcal{R}$} periodically \Comment{\cref{sec:probing}} \label{alg:line:probe}
		\State $V \gets \Call{ProgressValueEstim}{t, T, p(t), P}$ \Comment{\cref{sec:value-of-progress}} \label{alg:line:value}
		\For{each state $s \in \mathcal{R} \times \{\texttt{spot}, \texttt{od}, \texttt{idle}\}$}
		\State $\bar{L}_{s} \gets \Call{PredictLifetime}{s}$ \Comment{\cref{sec:duration-prediction}} \label{alg:line:predict}
		\State $U_{s} \gets \Call{CalcUtility}{s, V, \bar{L}_{s}}$ \Comment{\cref{sec:unified-cost}} \label{alg:line:utility}
		\EndFor
		\State Sort candidate states by $U_s$ descending
		\For{each $s=(r,m)$ with $U_s>U_{(r_0,m_0)}$} \Comment{\cref{sec:putting-together}}
        \State succeeds $\gets$ \texttt{\textbf{Launch}} $s \textbf{ or } m = \texttt{idle}$
		\If{succeeds}
		\State \texttt{\textbf{Terminate}} $(r_0,m_0)$ if $m_0\neq\texttt{idle}$
        \State $(r_0,m_0) \gets (r,m)$; \textbf{break}
		\EndIf
		\EndFor
		\EndWhile
        \State \texttt{\textbf{Terminate}} $(r_0,m_0)$ \Comment{Terminate on job completion}
	\end{algorithmic}
\end{algorithm}

\Cref{alg:policy} summarizes the scheduling policy. The core idea is simple: at each scheduling step, evaluate all candidates and migrate to any option that is better than the current state. The scheduler first applies the safety net when needed (\cref{sec:deadline-awareness}, lines~\ref{alg:line:safety}--\ref{alg:line:safety-end}). It then periodically sends lightweight probes to all regions (\cref{sec:probing}, line~\ref{alg:line:probe}). After computing the value of future progress based on the current deadline pressure (\cref{sec:value-of-progress}, line~\ref{alg:line:value}), the policy iterates over all candidate states $s = (r,m)$, predicts their lifetimes using the virtual instance view (\cref{sec:duration-prediction}, line~\ref{alg:line:predict}), and computes their utility $U_{(r,m)}$ (\cref{sec:unified-cost}, line~\ref{alg:line:utility}). It then compares each candidate against the current state's utility $U_{(r_0,m_0)}$.\footnote{In practice, we add a hysteresis threshold $\Delta$ to prevent thrashing: the scheduler switches only when $U_{(r,m)} > U_{(r_0,m_0)} + \Delta$.} Candidates are attempted in descending utility order, and the scheduler migrates to the first one that succeeds (idling always succeeds; spot may fail due to unavailability). This approach ensures the system continually moves toward better options as conditions change.

%% file: 5_implementation.tex
\section{\sys Implementation}

We implemented \sys, a general deadline-aware batch job execution system that combines spot and on-demand instances across multiple regions to guarantee deadline completion while minimizing cost. \sys is workload-agnostic, supporting any workload that periodically checkpoints to a persistent store, and it seamlessly manages checkpoint migration when moving across regions.
\sys is built on top of SkyPilot~\cite{skypilot}, an open-source multi-cloud system, and adds a scheduler on top of it with \textasciitilde{}6{,}000 lines of code. The implementation supports common training frameworks including PyTorch~\cite{pytorch-fsdp} and Hugging Face Transformers~\cite{huggingface}.

\MyPara{Scheduler.} The scheduler manages the entire lifecycle of the job. It first performs periodic probes to collect the data used in (\cref{sec:probing}). The probe interval is configurable and is set to 2 hours by default, which we find sufficient to capture key characteristics of the spot market while keeping probing overhead reasonable (\cref{sec:e2e}). The scheduler then forwards the probe results to the policy and provisions instances according to the policy's decision.

\MyPara{Checkpoint migration.} \sys assumes that the workload periodically checkpoints to a persistent store (e.g., a cloud bucket) and it handles checkpoint migration when a job is moved to another region. Because data on public clouds must be loaded onto an instance after it is provisioned, \sys implements a two-stage pipeline to accelerate migration. While the target instance is provisioning, \sys first copies the checkpoint to an object store in the target region. Once the instance becomes available, \sys downloads the checkpoint during runtime setup so that the job can resume execution promptly.

%% file: 7_evaluation.tex
\begin{figure*}[t]
	\centering
	\begin{subfigure}{0.33\textwidth}
		\includegraphics[width=\linewidth]{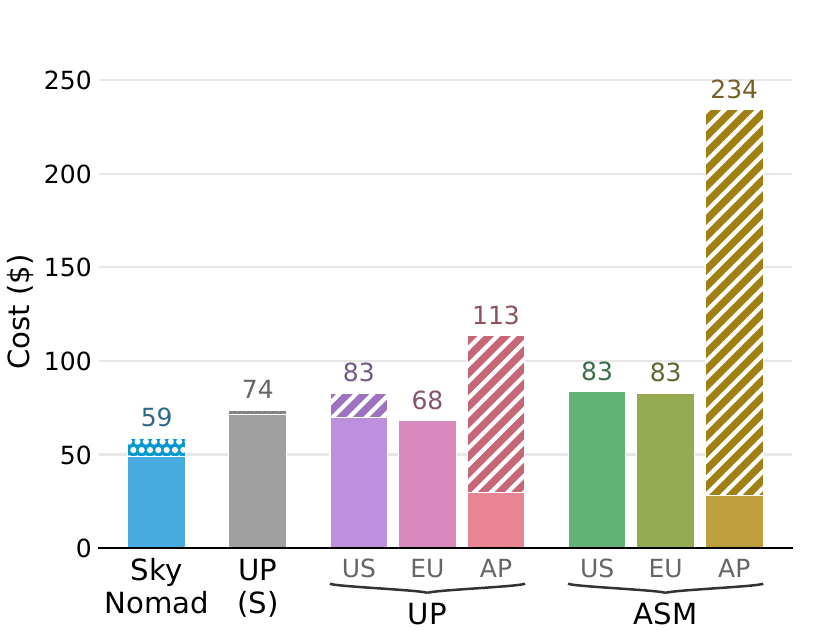}
		\caption{4$\times$L4 (\texttt{g6.12xlarge})}
		\label{fig:cost-comparison-L4}
	\end{subfigure}%
	\hfill
	\begin{subfigure}{0.33\textwidth}
		\includegraphics[width=\linewidth]{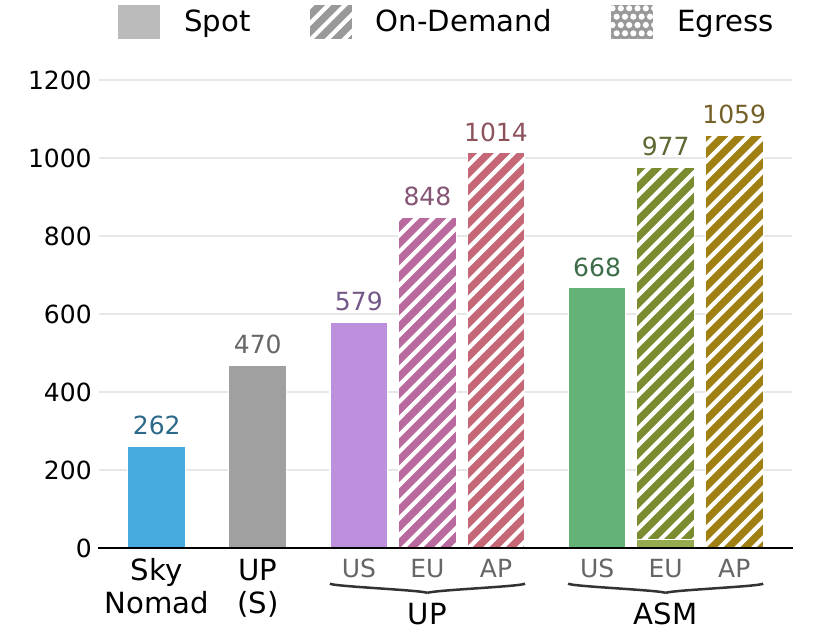}
		\caption{8$\times$A100 (\texttt{p4d.24xlarge})}
		\label{fig:cost-comparison-A100}
	\end{subfigure}%
	\hfill
	\begin{subfigure}{0.33\textwidth}
		\includegraphics[width=\linewidth]{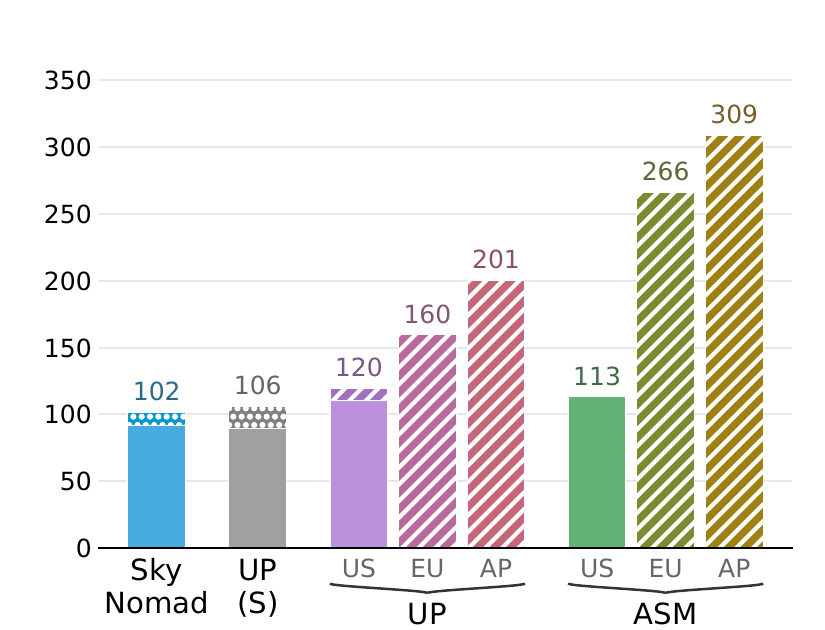}
		\caption{4$\times$A10G (\texttt{g5.12xlarge})}
		\label{fig:cost-comparison-A10G}
	\end{subfigure}
	\caption{End-to-end cost comparison across three accelerator types. \sys achieves the lowest cost in all configurations. Single-region baselines (UP and ASM) show high variance depending on regional spot availability. Only multi-region approaches incur egress costs (dotted region at bar tops). A100 shows minimal egress because the large checkpoint size (500\,GB) makes cross-region migration expensive, so the system prefers intra-region zone switches. All jobs complete within their deadlines.}
	\label{fig:cost-comparison}
\end{figure*}

\section{Evaluation}

We evaluate \sys comprehensively both on real cloud deployments that experience actual preemptions and in simulation using replayed spot traces. We seek to answer the following questions:

\begin{itemize}
	\item Can \sys reduce cost in real cloud environments while still completing jobs within their deadlines? (\cref{sec:e2e})
	\item Can \sys generalize across different accelerator types and their preemption patterns? (\cref{sec:e2e} and \cref{sec:gpus-selection})
	\item Can \sys achieve consistently low cost under varying deadline tightness, number of regions, and checkpoint sizes? (\cref{sec:deadline-sensitivity} to \cref{sec:checkpoint-sensitivity})
	\item Can \sys generalize across geographic deployments and data sovereignty constraints? (\cref{sec:region-selection})
\end{itemize}

\subsection{End-to-End Experiments}
\label{sec:e2e}

We run end-to-end experiments on AWS to fine-tune LLMs using three different accelerator configurations: 4$\times$L4, 8$\times$A100, and 4$\times$A10G. We launch all baseline systems and \sys \emph{simultaneously} so they experience the same real-time spot preemption events. The total cost of these experiments is approximately \$8,000.

\MyPara{Baselines.} We compare \sys with:

\begin{itemize}[leftmargin=10pt]
	\item \textbf{Uniform Progress (UP)~\cite{cant-be-late}}: A single-region deadline-aware policy that distributes progress evenly over time and uses on-demand and spot instances interchangeably to guarantee deadline completion.
	\item \textbf{AWS SageMaker Managed Spot (ASM)~\cite{sagemaker2025}}: A production-grade batch job execution system that uses spot instances to reduce cost. ASM relies exclusively on spot instances and can draw spot capacity from multiple availability zones within a single region, but only switches zones upon preemption and does not support multi-region.
	\item \textbf{UP(S)}: A multi-region extension of UP that we implemented as a baseline, representing SkyPilot's production failover policy~\cite{skypilot-failover}. Upon preemption, it \emph{Switches} the job to a \emph{different} region to avoid volatile periods~\cite{skyserve}, trying candidate regions sequentially from cheapest to most expensive until one succeeds. Between preemptions, it follows UP's progress distribution strategy to guarantee deadline completion and exploit rule~\cite{cant-be-late}, staying in the current region as long as it remains available. \footnote{We also design additional heuristics (UP(A), UP(AP)) to ablate individual components of \sys in \cref{sec:sim}.}
\end{itemize}

We run both UP and ASM separately in each of three regions (\texttt{us-west-2}, \texttt{eu-central-1}, and \texttt{ap-northeast-1}) to study how regional spot availability affects their performance. ASM is not inherently deadline-aware; to ensure a fair comparison, we manually trigger the safety net (\cref{sec:deadline-awareness}), switching to on-demand when needed to meet the deadline.

\MyPara{Workloads.} We use Hugging Face Transformers~\cite{huggingface} to fine-tune Qwen3 models~\cite{qwen3} on the Orca-Math dataset~\cite{orcamath}: Qwen3-4B on 4$\times$L4 (\texttt{g6.12xlarge}) and 4$\times$A10G (\texttt{g5.12xlarge}), and Qwen3-14B on 8$\times$A100 (\texttt{p4d.24xlarge}). Each job runs on a single instance and requires about 30 hours of compute with a 45-hour deadline. Checkpoint sizes are 100\,GB and 500\,GB, respectively. Cold start time, including instance provisioning, environment setup, and checkpoint loading, is approximately 6 minutes.

\MyPara{Results.} \cref{fig:cost-comparison} shows the total cost for each system across the three accelerator configurations. \sys achieves the lowest cost in all cases: 10--55\% savings versus the best single-region baseline, 47--69\% versus the average single-region baseline, and 4--44\% versus the multi-region baseline UP(S). These savings already account for \sys’s overheads: egress costs for cross-region migrations account for up to 16\% of total cost (dotted portion in \cref{fig:cost-comparison}), while probing costs are negligible at \$1--3 per job. Despite the egress overhead, cross-region migration enables \sys to exploit cheaper spot capacity that would otherwise be inaccessible.

\MyPara{Single region solutions are inherently limited.} Single-region baselines face two limitations. First, the ever-changing nature of the spot market means no single region consistently offers the best availability or price. A region that performs well at the start of a job may become unavailable mid-execution, and static region selection cannot adapt to such changes. In our experiments across three representative regions, UP's total cost varies by up to 1.7$\times$ on L4~(\cref{fig:cost-comparison-L4}) and 1.8$\times$ on A100~(\cref{fig:cost-comparison-A100}). While the best region \texttt{eu-central-1} for L4~(\cref{fig:cost-comparison-L4}) achieves low cost with abundant spot capacity, the worst-performing region (\texttt{ap-northeast-1}) has no spot available for more than 70\% of the time and must fall back to expensive on-demand instances, driving 74\% of its total cost. ASM shows even higher variance (up to 2.8$\times$ on A10G) because its zone-level failover may land on a more expensive zone~(\cref{fig:cost-comparison-A10G}).
Second, spot availability patterns differ across accelerator types: \texttt{eu-central-1} performs well for L4 but poorly for A100, while \texttt{us-west-2} shows the opposite pattern. As a result, region choices informed by one GPU type do not necessarily transfer to another.
Given these limitations, even the best-performing single-region baseline (UP in \texttt{eu-central-1}) costs 10\% more than \sys.

\begin{figure}[t]
	\centering
	\includegraphics[width=\linewidth]{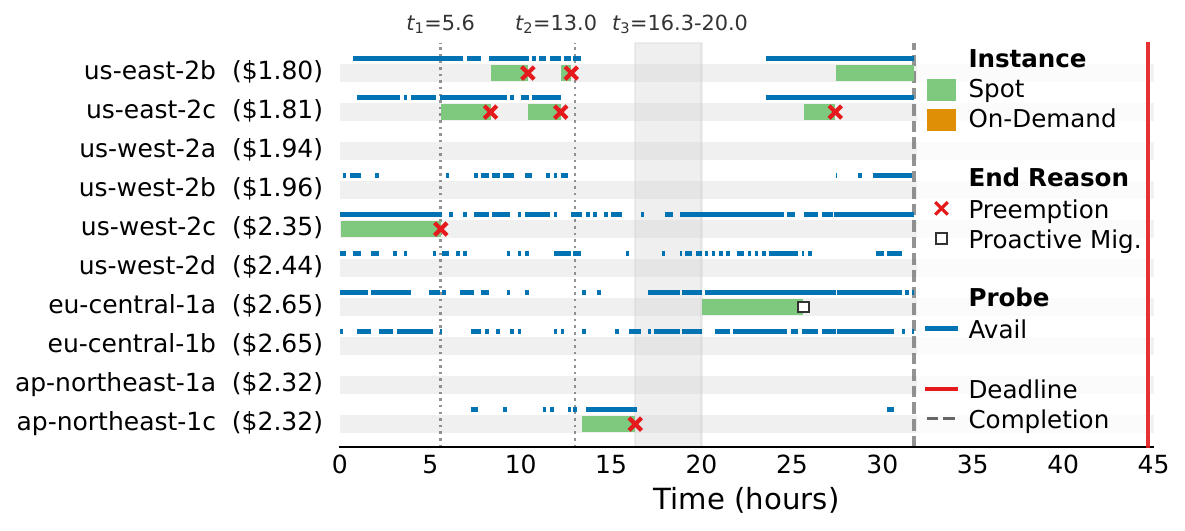}
	\caption{Migration traces for \sys in the L4 experiment. Green bars indicate spot instance usage; blue segments show probe-detected availability; $\square$ markers indicate proactive migration decisions.}
	\label{fig:migration-trace}
\end{figure}

\MyPara{Naive multi-region is not enough.} Multi-region approaches achieve lower and more stable costs by drawing spot capacity from multiple regions; both \sys and UP(S) benefit from the aggregated spot resources across regions and complete the job without using any on-demand instances (\cref{fig:cost-comparison-L4}).
However, UP(S) only migrates when preempted. If it lands in an expensive region that happens to have stable availability, it stays there indefinitely. As a result, despite having access to all three regions, UP(S) underperforms the best single-region baseline on L4~(\cref{fig:cost-comparison-L4}).
\sys, in contrast, continuously monitors all candidate zones through probing and proactively migrates when expected savings outweigh migration costs. This enables \sys to save 44\% over UP(S) on A100~(\cref{fig:cost-comparison-A100}). For example, in \cref{fig:migration-trace}, continuous probing of the cheaper zone \texttt{us-east-2b/c} (\$1.8/hr) detects an availability window at hours 23--25 and predicts a lifetime of 3.5 hours. Under the unified cost model, this yields a higher utility than \texttt{eu-central-1a}, triggering a price-aware proactive migration ($\square$ marker). Such real-time visibility distinguishes \sys from reactive policies like UP(S).

\MyPara{\sys sees the whole picture.} Unlike heuristics that consider only one metric (e.g., price or availability), \sys jointly considers price, predicted lifetime, egress cost, and deadline pressure. We highlight three decisions in the L4 trace (\cref{fig:migration-trace}) that illustrate this.

\noindent\textbf{(1) $t_1 = 5.6$\,hr:} After preemption in \texttt{us-west-2c}, \sys evaluates all available regions. Probing shows \texttt{us-east-2c} (\$1.81/hr) is available with predicted lifetime of 4 hours. At this point, $V(t) \approx \$2.6$/hr, and after accounting for cold start overhead (6 minutes), each hour of effective progress is worth \$2.5. The \$2 egress cost, amortized over the 4-hour lifetime, adds \$0.50/hr. The combined per hour cost is \$1.81 + \$0.50 = \$2.31/hr, which is lower than the monetary value of progress, yielding positive utility and \sys thus migrates.

\noindent\textbf{(2) $t_2 = 13.0$\,hr:} After 4 preemptions in \texttt{us-east-2} (hours 9--13), the job has fallen behind schedule with deadline pressure $\theta(t) = 0.73$ (vs.\ nominal $P/T = 0.67$; see \cref{sec:value-of-progress}), raising $V(t)$ to $\approx \$4$/hr. At this elevated value of progress, even the more expensive \texttt{ap-northeast-1c} (\$2.32/hr) yields positive utility. Meanwhile, \texttt{us-east-2} is detected to be in a volatile period ($\gamma^* > 1$), reducing its predicted lifetime to under 1 hour. \sys then selects \texttt{ap-northeast-1c} as the only region with positive utility and tolerates the \$2 egress cost.

\noindent\textbf{(3) $t_3 = 16.3$--$20.0$\,hr:} After a spot preemption in region \texttt{ap-northeast-1c}, \sys attempts to launch in the cheap \texttt{us-east-2} regions (\$1.80/hr) first (highest utility), but they are unavailable. Other available regions like \texttt{us-west-2c} (\$2.35/hr) appear volatile and yield near-zero or negative utility given $V(t) \approx \$2.6$/hr, while idling has $U = 0$. \sys waits, and as deadline pressure increases, $V(t)$ rises until \texttt{eu-central-1a} (\$2.65/hr) finally yields positive utility at hour 20, at which point \sys launches there.

\MyPara{Summary.} End-to-end experiments confirm that \sys meets all deadlines while achieving 55\% cost savings on average over single-region baselines (up to 70\%) through its unified cost model, which balances price, availability, migration cost, and deadline pressure.

\subsection{Simulation with Spot Availability Trace}
\label{sec:sim}

To understand whether \sys can perform consistently well under diverse conditions, we replay real spot availability traces in simulation. This allows us to systematically study how parameters such as deadline tightness, number of regions, and checkpoint sizes affect the policy at a large scale.

\subsubsection{Experimental Setup}
\label{sec:sim-setup}

\MyPara{Traces.} We collect spot availability traces for 16 1$\times$H100 instances (\texttt{a3-highgpu-1g}) across 13 GCP zones over a 14-day period by probing each zone every 10 minutes, at a total cost of approximately \$9{,}000. We also use publicly available traces for 1$\times$V100 instances on AWS~\cite{cant-be-late} for generalizability.

\MyPara{Baselines.} We compare \sys with:
\begin{itemize}[leftmargin=10pt]
	\item \textbf{Optimal}: An omniscient policy with full knowledge of future spot availability. We use dynamic programming to compute its optimal decisions, which serve as a lower bound on achievable cost while completing the job.
	\item \textbf{\sys(o)}: A variant of \sys with an oracle for the next spot lifetime in each region.
	\item \textbf{UP}: Single-region Uniform Progress~\cite{cant-be-late}, same as in \cref{sec:e2e}. We report the average cost across all regions.
	\item \textbf{UP(S)}: Same as in \cref{sec:e2e}.
	\item \textbf{UP(A)}: A multi-region extension of UP that uses the same probing mechanism as \sys (\cref{sec:probing}) and selects regions based on observed availability, defined as the fraction of successful probes in a sliding window of the last 5 samples.
	\item \textbf{UP(AP)}: A multi-region extension of UP that selects regions based on availability divided by spot price, balancing availability against cost without lifetime prediction.
\end{itemize}

\MyPara{Default parameters.} Unless otherwise specified, we use a job duration of 100 hours with a 150-hour deadline, a checkpoint size of 50\,GB, and a cold start overhead of 6 minutes. We use 8 regions by default.

\begin{figure}[t]
	\centering
	\begin{subfigure}[t]{0.48\columnwidth}
		\centering
		\includegraphics[width=\linewidth]{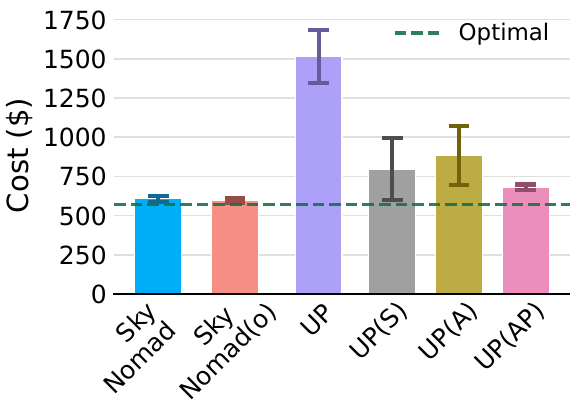}
		\caption{H100}
		\label{fig:sim-different-gpus-h100}
	\end{subfigure}
	\begin{subfigure}[t]{0.48\columnwidth}
		\centering
		\includegraphics[width=\linewidth]{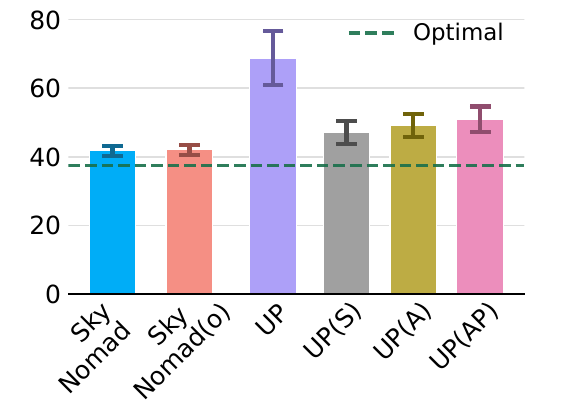}
		\caption{V100}
		\label{fig:sim-different-gpus-v100}
	\end{subfigure}
	\caption{Cost on two independent spot traces: (a) H100 on GCP and (b) V100 on AWS. The green horizontal line indicates the Optimal policy cost. \sys generalizes across cloud providers and GPU types. We simulate 20 jobs with different start times; error bars show standard error.}
	\label{fig:sim-different-gpus}
\end{figure}
\subsubsection{Generalizability across Accelerators}
\label{sec:gpus-selection}

\begin{figure}[t]
	\centering
	\includegraphics[width=\linewidth]{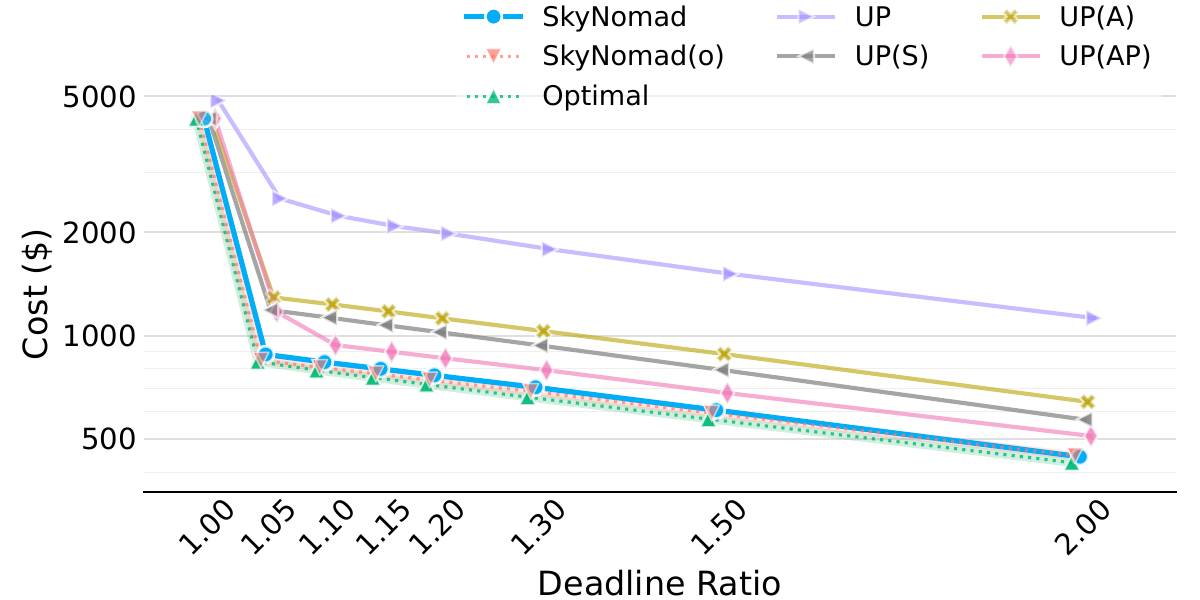}
	\caption{Cost with respect to varying deadline ratios. The y-axis is in log scale. A ratio of 1.0 is a theoretical lower bound.\footnotemark}
	\label{fig:deadline-sensitivity}
\end{figure}
\addtocounter{footnote}{-1}
\stepcounter{footnote}\footnotetext{In practice, cold start overhead requires a slightly higher ratio.}

\Cref{fig:sim-different-gpus} compares policy performance on two independent spot traces: our H100 trace from GCP and a publicly available V100 trace from AWS~\cite{cant-be-late}.

\sys achieves near-optimal cost on both traces: \$610 on H100 (within 11\% of Optimal) and \$42 on V100 (within 12\% of Optimal). This demonstrates that our approach generalizes across cloud providers and GPU types.

We define \emph{selection accuracy} as the fraction of time a policy uses the cheapest available region, capturing its ability to identify and leverage the best option throughout execution. \sys achieves 83--100\% selection accuracy across both traces. In contrast, UP(A) never selects the cheapest region, yielding 0\% selection accuracy on both V100 and H100: it always chooses the most available region regardless of price. This explains the cost gap: the unified cost model's price term (\cref{sec:unified-cost}) steers selection toward cost-effective regions, while availability-only heuristics ignore price entirely.

The gap between \sys and \sys~(o) is under 5\% on both traces, with 95--99\% region selection overlap, indicating that \sys's lifetime prediction leads to nearly identical migration decisions as the oracle. Since \sys~(o) uses an oracle for spot lifetime, this small gap confirms that our survival-analysis-based lifetime prediction (\cref{sec:duration-prediction}) estimates lifetimes accurately enough to make near-optimal decisions. On V100, \sys slightly outperforms \sys(o); this is within the variance of the simulation.

\subsubsection{Impact of Deadline Tightness}
\label{sec:deadline-sensitivity}

\Cref{fig:deadline-sensitivity} shows how cost varies with deadline tightness. The deadline ratio is defined as $T/P$, where $T$ is the deadline and $P$ is the job duration. Larger ratios provide more flexibility to wait for spot capacity.

At a deadline ratio of 1.0, all policies immediately trigger the safety net (\cref{sec:deadline-awareness}) and run entirely on on-demand instances, converging to \$4{,}000--\$5{,}000. As slack increases, \sys tracks Optimal closely, reducing cost to \$420 at 2.0$\times$ and staying within 10\% of Optimal across all tested ratios.

Single-region UP scales poorly, remaining at \$1{,}100 (2.6$\times$ \sys) even at 2.0$\times$ slack. Multi-region policies scale better by drawing capacity from multiple regions, but still lag \sys by 15--30\%. The lag is not due to on-demand fallback, as multi-region policies rarely trigger the safety net. Instead, they select expensive regions: UP(A) greedily chooses high-availability regions regardless of price (up to 3.6$\times$ cost in Asia), while UP(AP) and UP(S) react to preemptions without considering predicted lifetime.

\begin{figure}[t]
	\centering
	\includegraphics[width=\linewidth]{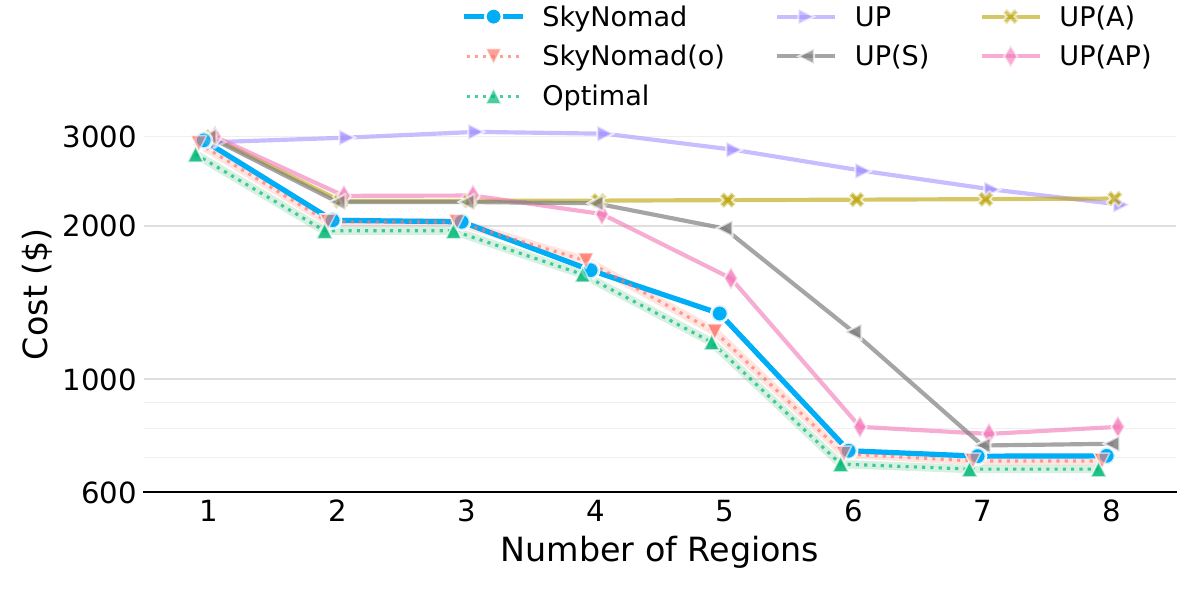}
	\caption{Cost with respect to varying number of available regions. For all systems, cost decreases monotonically when more regions are available. In all cases, \sys achieves near-optimal cost.}
	\label{fig:region-scaling}
\end{figure}

\begin{figure}[t]
	\centering
	\includegraphics[width=\linewidth]{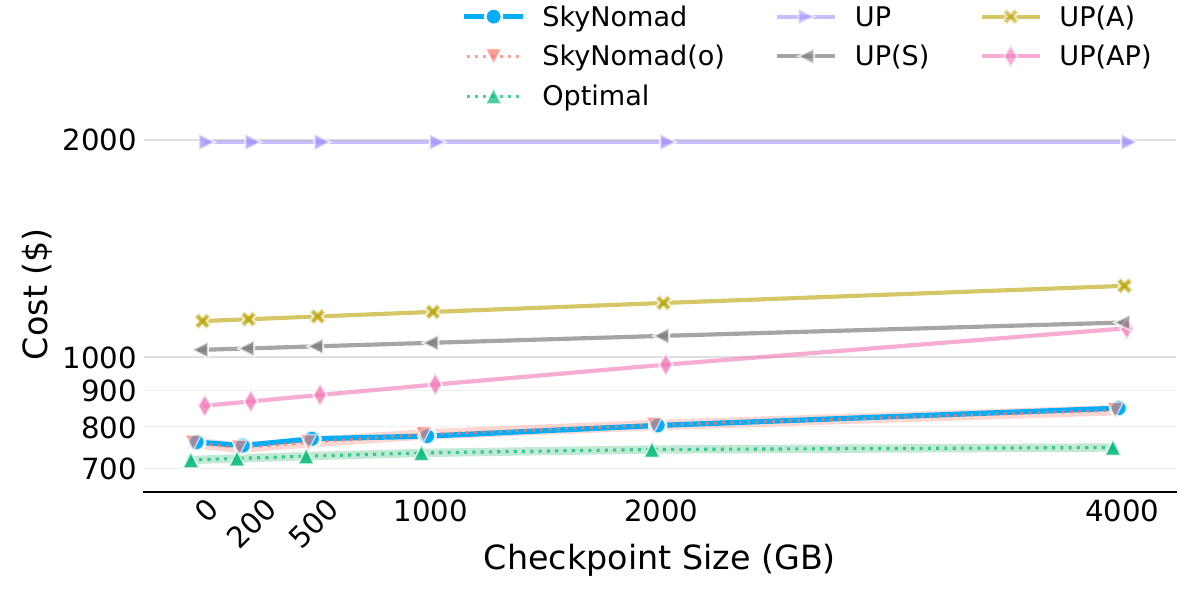}
	\caption{Cost with respect to varying checkpoint sizes. Larger checkpoints increase migration cost, reducing the benefit of cross-region scheduling. \sys scales gracefully by amortizing migration cost over predicted lifetimes, while heuristics like UP(AP) suffer from frequent migrations.}
	\label{fig:checkpoint-sensitivity}
\end{figure}

\subsubsection{Impact of Number of Regions}
\label{sec:region-scaling}

\Cref{fig:region-scaling} shows how cost changes as we vary the number of candidate regions from 1 to 8. We incrementally add regions in order of their average spot availability.

With one region, all policies perform similarly (\$2{,}800--\$3{,}000) since there is no region selection to optimize. As regions increase, \sys approaches Optimal: at 8 regions, \sys achieves \$707, within 6\% of Optimal (\$666) and 2\% of \sys(o) (\$691).

Naive heuristics plateau or underperform. \sys achieves 95\% region selection overlap with Optimal, while UP(A) achieves only 0.2\%. UP(A) settles in an available but expensive region and never migrates away: in the 8-region scenario, UP(A) selects \texttt{asia-south2-b} (95\% availability, but 4$\times$ the cheapest price) 94\% of the time, resulting in 4.8$\times$ higher cost than Optimal. UP(S) considers only price and ignores availability, generally underperforming UP(AP). UP(AP) tracks \sys more closely (\$806 at 8 regions) by balancing availability and price, but cannot match our unified cost model.

Beyond 6 regions, returns diminish: \sys and Optimal improve steadily up to 6 regions, after which aggregated availability saturates and new regions contribute overlapping availability windows rather than complementary capacity. For users with geographic constraints (\cref{sec:region-selection}), even 2--3 well-chosen regions suffice for significant savings.

\subsubsection{Impact of Checkpoint Size}
\label{sec:checkpoint-sensitivity}

Checkpoint size affects migration cost since egress cost is proportional to the amount of data transferred. \Cref{fig:checkpoint-sensitivity} varies checkpoint size from 0 to 4\,TB to cover workloads ranging from batch inference with small index state to large-scale training (full model and optimizer state).

Single-region UP is unaffected by checkpoint size since it never migrates across regions. \sys scales gracefully: at 4\,TB, \sys reaches \$850, 57\% cheaper than UP at \$2{,}000, by reducing migration frequency 70\% compared to UP(AP)---preferring to wait for spot recovery rather than incur expensive cross-region transfers. The slight cost decrease from 200\,GB to 500\,GB for \sys is within simulation variance.

Heuristics that react to frequently changing signals suffer at large checkpoints. UP(AP) exhibits the steepest cost growth: availability shifts as new probes arrive, and dividing by price amplifies these fluctuations in cheap regions, so a small availability change can flip the ratio ranking and trigger a region switch. UP(A) is more stable because high-availability regions tend to remain high. UP(S) is most stable since prices rarely change, so it returns to the same cheap regions.

\subsubsection{Impact of Data Sovereignty Requirements}
\label{sec:region-selection}

\begin{figure}[t]
	\centering
	\includegraphics[width=\columnwidth]{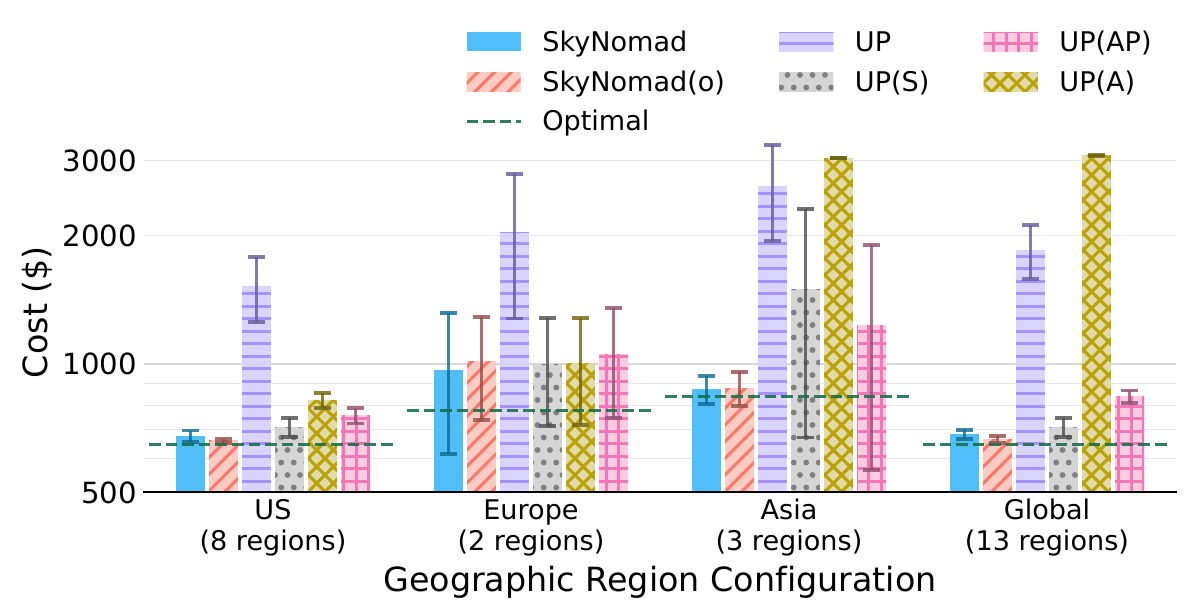}
	\caption{Cost under geographic constraints (US-only, Europe-only, Asia-only, and Global). \sys adapts effectively across all configurations, achieving significant savings even with limited regional diversity. Error bars show 95\% confidence intervals over 20 jobs.}
	\label{fig:region-selection}
\end{figure}

Data sovereignty requirements (e.g., GDPR~\cite{gdpr-cloud}) may restrict jobs to specific geographic regions to prevent data from leaving a designated area. We simulate such constraints by limiting candidate regions to the same continent. \Cref{fig:region-selection} evaluates \sys under four configurations: US-only (3 regions), Europe-only (2 regions)\footnote{Due to quota limits, we only have access to two European regions.}, Asia-only (3 regions), and Global (all 13 regions).
\sys performs well even with geographic constraints: \$700 in the US (3 regions), close to the Global optimum of \$650; \$950 in Europe (2 regions), still 52\% cheaper than UP; and \$870 in Asia (3 regions), 68\% cheaper than UP. Even with only 2--3 regions, \sys achieves significant savings, confirming that a modest number of well-chosen regions suffices (\cref{sec:region-scaling}). Asia shows the highest variance due to heterogeneous availability patterns. UP(A) performs particularly poorly (\$3{,}038, 3.6$\times$ Optimal) by selecting the most expensive region (\texttt{asia-south2-b}) 100\% of the time, achieving 0\% selection accuracy. In contrast, \sys achieves 100\% selection accuracy by consistently selecting the cheapest region (\texttt{asia-southeast1-b}), reaching cost within 0.1\% of Optimal.

%% file: 9_related.tex
\section{Related Work}
\label{related-work}

\MyPara{Spot instances for training and batch jobs.}
Spot instances have been widely studied for reducing the cost of batch computations, including HPC~\cite{hpdc-spot, yi2012monetary}, analytics~\cite{tr-spark, flint}, and ML training~\cite{deepspotcloud, bamboo, varuna, oobleck, parcae}.
For example, Bamboo~\cite{bamboo} provides resilience through redundant computation, Varuna~\cite{varuna} dynamically morphs training configurations, and Parcae~\cite{parcae} predicts spot availability and live-migrates tasks.
However, these systems aim to maximize throughput and provide no guarantees on meeting deadlines.
They rely exclusively on spot instances, so if spot capacity disappears for extended periods (which can last days~\cite{skyserve}), progress halts and deadlines cannot be guaranteed.

\MyPara{Deadline-aware spot scheduling.}
Uniform Progress (UP)~\cite{cant-be-late} uses spot when available and falls back to on-demand as deadlines approach, achieving 27--84\% savings over on-demand only.
However, UP operates in a single region and cannot exploit multi-region spot capacity: when one region has no spot available, another may still have capacity.
Proteus~\cite{proteus} proposed tiered reliability with transient servers, but operates within a single region.
Snape~\cite{snape} uses RL to distribute capacity across spot and on-demand VMs, but targets long-running services with SLO requirements rather than batch jobs with deadlines.
\sys is the first to combine deadline guarantees with multi-region spot scheduling, exploiting spatial and temporal heterogeneity to minimize cost.

\MyPara{Spot instances for serving.}
Systems like MArk~\cite{mark}, Cocktail~\cite{cocktail}, and Tributary~\cite{tributary} use spot for inference serving with response-time SLOs.
SpotServe~\cite{spotserve} parallelizes LLM inference across spot GPUs and adapts to preemptions, while SkyServe~\cite{skyserve} replicates instances across regions to maintain availability.
However, serving systems cannot exploit slack: they must provision on-demand immediately when spot disappears to maintain latency SLOs.
Batch jobs can instead wait for spot capacity to return, enabling more cost reduction without sacrificing deadline requirements.

%% file: 8_discussion.tex
\section{Discussion}
\label{sec:discussion}

\MyPara{Multi-Cloud Extension.}
\sys currently operates within a single cloud provider. Extending to multi-cloud could further expand the spot capacity pool, but introduces challenges including heterogeneous APIs, higher cross-cloud migration costs, and provider-specific probing semantics. We leave this as future work.

\MyPara{Availability Signals.}
We use lightweight probing as our availability signal, and it incurs only a modest cost (\$1–3 per job). Cloud providers are beginning to offer native availability signals (e.g., AWS Spot Placement Score~\cite{aws-spot-placement-score}) that could reduce probing cost but at the expense of portability.

\MyPara{Autoscaling Integration.}
\label{sec:autoscaling}
\sys uses a fixed number of instance with gang-scheduled preemption for compatibility. Combining multi-region scheduling with elastic autoscaling~\cite{bamboo,spotserve,varuna,oobleck} could yield additional savings, but requires workload-specific support for dynamic parallelism.

\MyPara{Progress Loss After Preemptions.}
Computation after the last checkpoint is lost upon preemption. \sys's lifetime prediction could inform adaptive checkpointing~\cite{checkfreq,gupta2024just}, where the system checkpoints more frequently when the remaining lifetime is predicted to be short. We leave this as future work.

\MyPara{Task-as-a-Service Pricing Model.}
Single-region scheduling suffers from high cost variance. Multi-region scheduling provides more consistent costs through geographic diversity, potentially enabling task-as-a-service pricing models that abstract away spot market volatility.

%% file: 10_conclusion.tex
\section{Conclusion}

\sys shows that exploiting spatial and temporal spot heterogeneity can greatly reduce AI batch job cost still meeting deadlines. With real-time probing, lifetime prediction, and a unified cost model that captures deadline pressure, and operational cost, \sys continually selects the most cost-effective region and migrates when beneficial. In real-world deployments, \sys reduces costs by 1.25--3.96$\times$ compared to existing research and production systems.